\documentclass[useAMS,usenatbib]{mnras}

\usepackage{mathptmx}
\usepackage{ amssymb }
\usepackage{multirow}
\usepackage{psfrag}
\usepackage{pspicture}
\usepackage{graphicx}
\usepackage{amsmath}
\usepackage{color}
\usepackage{url}
\usepackage{caption}

\usepackage{subcaption}
\usepackage{grffile}
\usepackage{multirow}
\usepackage{multicol}
\usepackage{dcolumn}
\usepackage{hyperref}
 
\usepackage{setspace}

\title[H$\alpha$ survey of galaxy clusters]{A large H$\alpha$ survey of star formation in relaxed and merging galaxy cluster environments at $z\sim0.15-0.3$}
\author[A. Stroe et al.]{Andra Stroe$^{1}$\thanks{E-mail: astroe@eso.org}\thanks{ESO Fellow}, David Sobral$^{2,3}$\thanks{VENI Fellow}, Ana Afonso$^{4,5}$, Lara Alegre$^{4,5}$, Jo\~ao Calhau$^{2}$,
\newauthor Sergio Santos$^{2}$, Reinout van Weeren$^6$\\
$^{1}$European Southern Observatory, Karl-Schwarzschild-Str. 2, 85748, Garching, Germany\\
$^{2}$Department of Physics, Lancaster University, Lancaster, LA1 4YB, UK\\
$^{3}$Leiden Observatory, Leiden University, P.O.\ Box 9513, NL-2300 RA Leiden, The Netherlands\\
$^{4}$Instituto de Astrof\'{\i}sica e Ci\^{e}ncias do Espa\c{c}o, Universidade de Lisboa, OAL, Tapada da Ajuda, PT1349-018 Lisbon, Portugal \\
$^{5}$Departamento de F\'{i}sica, Faculdade de Ci\^{e}ncias, Universidade de Lisboa, Edif\'{i}cio C8, Campo Grande, PT1749-016 Lisbon, Portugal \\
$^{6}$Harvard Smithsonian Center for Astrophysics, 60 Garden Street, MS-06
Cambridge, MA 02138\\
}
\begin{document}
\maketitle
\begin{abstract}
We present the first results from the largest H$\alpha$ survey of star formation and AGN activity in galaxy clusters. Using 9 different narrow band filters, we select $>3000$ H$\alpha$ emitters within $19$ clusters and their larger scale environment over a total volume of $1.3\times10^5$ Mpc$^3$. The sample includes both relaxed and merging clusters, covering the $0.15-0.31$ redshift range and spanning from $5\times10^{14}$ $M_{\odot}$ to $30\times10^{14}$ $M_{\odot}$. We find that the H$\alpha$ luminosity function (LF) for merging clusters has a higher characteristic density $\phi^*$ compared to relaxed clusters. $\phi^*$ drops from cluster core to cluster outskirts for both merging and relaxed clusters, with the merging cluster values $\sim0.3$ dex higher at each projected  radius. The characteristic luminosity $L^*$ drops over the $0.5-2.0$ Mpc distance from the cluster centre for merging clusters and increases for relaxed objects. Among disturbed objects, clusters hosting large-scale shock waves (traced by radio relics) are overdense in H$\alpha$ emitters compared to those with turbulence in their intra-cluster medium (traced by radio haloes). We speculate that the increase in star formation activity in disturbed, young, massive galaxy clusters can be triggered by interactions between gas-rich galaxies, shocks and/or the intra-cluster medium, as well as accretion of filaments and galaxy groups. Our results indicate that disturbed clusters represent vastly different environments for galaxy evolution compared to relaxed clusters or average field environments.
\end{abstract}
\begin{keywords}
galaxies: luminosity function, mass function, galaxies: evolution, galaxies: formation, galaxies: clusters: general, cosmology: large-scale structure of Universe
\end{keywords}

\section{Introduction}\label{sec:intro}

Since the dawn of the first stars and the first galaxies up to the present age, there has been tremendous evolution in galaxy populations \citep[e.g.][]{1996MNRAS.283.1388M,1996ApJ...460L...1L,2006ApJ...651..142H, 2014ARA&A..52..415M}. Star formation (SF) activity steadily rose up to $z\sim2-3$, but has been declining since then \citep{1996ApJ...460L...1L, 2011ApJ...730...61K,2013MNRAS.428.1128S, 2015MNRAS.453..242S}. This evolution is reflected in the properties of star-forming galaxies: the typical star formation rate (SFR) of galaxies (SFR$^*$) at $z\sim2$ is a factor $\sim10$ higher than in the local Universe \citep[e.g.][]{2013MNRAS.428.1128S,2014MNRAS.437.3516S}, while the specific star formation rate (sSFR) of galaxies at fixed mass increases with redshift by approximately the same amount \citep[e.g.][]{2012ApJ...757L..22F, 2013MNRAS.434..423K, 2014MNRAS.437.3516S}. Half of the stellar mass observed today was formed before $z\sim1$, when the Universe was about a third of its current age \citep[e.g.][]{2009ApJ...701.1765M, 2013ApJ...777...18M,2014ARA&A..52..415M}.

The properties of galaxies do not only vary with cosmic time, but also with environment \citep[e.g.][]{2010ApJ...721..193P, 2012ApJ...757....4P, 2016ApJ...825..113D}. There is a strong correlation between local density and the properties of the galaxy population. At $z<1$, massive elliptical galaxies are located at the centres of virialised clusters. Additionally, the general galaxy population in these clusters is dominated by passive, ellipticals and S0s \citep{1980ApJ...236..351D,1980ApJS...42..565D, 1997ApJ...490..577D}. The fraction of star-forming galaxies increases with radius from the cluster centre towards the cluster outskirts. The star-forming fraction is even higher in the large scale array of filaments surrounding clusters and in properly isolated field galaxies \citep{1980ApJ...236..351D}. Typical cluster environments prevent formation of new stars, either by maintaining galaxies quenched or by accelerating quenching processes \citep[e.g.][]{1978ApJ...219...18B, 1978ApJ...226..559B, 1980ApJ...236..351D}. Environmental quenching is so effective that, at low redshifts ($z<0.1$), the fraction of star-forming galaxies within relaxed clusters is below that in blank fields as far as three times the virial radius of the clusters \citep{2011ApJ...743...34C}. Therefore, despite the high density of galaxies within clusters, the number density of star-forming galaxies is lower in clusters than in average fields \citep[e.g.][]{1980ApJ...236..351D, 2003MNRAS.346..601G}. The potential transformation of field spirals into cluster ellipticals and S0s has been attributed to a number of processes: ram pressure stripping of the gas content infalling galaxies by the intracluster medium \citep[ICM, e.g.][]{1972ApJ...176....1G, 2014MNRAS.445.4335F}, gas removal \citep[strangulation,][]{1980ApJ...237..692L} and truncation of the halo and disk \citep[harassment,][]{1996Natur.379..613M} by tidal forces caused by interactions with other cluster galaxies or by gradients in the cluster gravitational potential. 

So far, most studies have focused on field galaxies or on galaxies in relaxed clusters. However, less literature has been dedicated to intermediate-density environments, such as filaments, and non-relaxed clusters, which provide a very different environment for the galaxies to interact with, compared to relaxed clusters. Filamentary structures and the outskirts of merging clusters host shock waves with Mach numbers between $\sim3$ and $\sim10$ \citep{2006MNRAS.367..113P, 2011MNRAS.418..960V, 2016MNRAS.458.2080B}, while the more central areas of merging clusters have increased turbulence. Recent studies indicate that non-relaxed clusters might display a reversal of the typical relaxed cluster environmental trends \citep{2014MNRAS.438.1377S, Stroe2015}. For example, star-forming tails and H$\alpha$ emitting galaxies were found near the shocks in the clusters Abell 2744 \citep{2012ApJ...750L..23O} and Abell 521 \citep{2003A&A...399..813F, 2004ApJ...601..805U}. Abell 2384 hosts an unexpected population of disk galaxies towards the cluster core  \citep{2014A&A...570A..40P}. Similarly, \citet{2004A&A...416..839B} find a significant population of active galaxies in the dynamically young cluster Abell 2219. \citet{2014ApJ...796...51D} find a higher fraction of H$\alpha$ emitting galaxies in filaments than in other environments. These galaxies are more metal rich and have lower inter-stellar medium electron densities than their field counterparts \citep{2015ApJ...814...84D}. The young massive merging cluster CIZA J2242.8+5301 \citep[`Sausage' cluster,][]{2007ApJ...662..224K} was found to host a large population of star-forming galaxies and AGN with high SFR, increased metallicity, lower electron densities (similar to filaments) and winds \citep{2014MNRAS.438.1377S, Stroe2015, 2015MNRAS.450..630S}. The similarly massive 1RXS J0603.3+4214 cluster \citep[`Toothbrush',][]{2012A&A...546A.124V} was found to be devoid of star-forming galaxies, an effect which may be attributed to the longer period passed since the subclusters merged \citep[$2$\,Gyr for the `Toothbrush' compared to $<1$\,Gyr for the `Sausage';][]{Stroe2015}. 

A range of SF tracers can be used to track the continuous transformation of galaxies across cosmic time and environment \citep[e.g.][]{2014ARA&A..52..415M}. However, different tracers are sensitive to different time scales, leading to different selection functions. Comparing studies performed with different SF tracers can result in contradicting conclusions regarding the SF evolution with cosmic time and environment. Many surveys of both clusters and fields \citep[e.g.][]{1999ApJ...527...54B, 2004MNRAS.348.1355B, 2007ApJS..172...70L, 2011A&A...529A.128B, 2014ApJ...783..136C, 2015A&A...576A..79L} use deep spectroscopy to study the SF properties of galaxies selected based on broad band (BB) photometry. Such surveys provide unique insight into the detailed physical processes of the surveyed galaxies. However, spectroscopic surveys have complicated selection functions, which, in many cases, do not only depend on the mass or SFR of the galaxies, but suffer from constrains in placement of fibres/slits. Achieving spectroscopic completeness is particularly difficult for clusters of galaxies, where the density of sources is very high and taking a spectrum for each galaxy requires numerous pointings with different fibre/slit placements. Candidate cluster members are most easily selected for spectroscopic follow-up through the red-sequence method, which ensures the galaxies are selected around the right redshift range. However, this method is biased against star-forming galaxies, selecting, by design, passive galaxies. Therefore, one of the main challenges is to obtain comparable samples of star-forming galaxies at different redshifts and in a range of environments, uniformly selected down to the same SFR limit. 

An efficient technique to uniformly select galaxies undergoing recent SF (averaged over $\sim10-20$ Myr) is to use the narrow-band (NB) technique to trace H$\alpha$ emission within a small redshift range \citep[e.g.][]{1995MNRAS.273..513B}. A NB filter which captures H$\alpha$ emission as well as the stellar continuum is used in combination with a BB filter which is dominated by stellar continuum. By subtracting the BB from the NB, emission line systems can be easily uncovered. This technique is ideal for selecting field star-forming galaxies at many different narrow redshift slices within which not much evolution is expected. The NB technique is also very well suited for identifying emission-line systems in clusters, ensuring selection of all cluster members within the plane of the sky as well as in the redshift direction \citep[e.g.][]{2002A&A...384..383I, 2004MNRAS.354.1103K, 2011MNRAS.416.2041M, 2011MNRAS.411..675S, 2013MNRAS.434..423K,2014MNRAS.438.1377S}.

As mentioned before, violent merging clusters and filamentary environments are expected to lead to a different evolution for galaxies than relaxed clusters. It is therefore important to quantify the nature and evolution of galaxies in the largely unexplored parameter space of merging and relaxed clusters as well as the cosmic web around them. These low and mid redshift ($z\sim0.1-0.7$) disrupted environments might be very similar to high-redshift ($z\sim1-5$) clusters and protoclusters, and can therefore serve as ideal counterparts to easily study. Pilot analyses of the `Sausage' and `Toothbrush' merging clusters \citep{2014MNRAS.438.1377S, Stroe2015, 2015MNRAS.450..630S} indicate that shocks in young mergers may induce SF in merging cluster galaxies. Could the turbulence also lead to enhanced SF? Could the different merger histories of clusters play a significant role? What is the dependence of SF on the mass of the host cluster? Is the merging activity more important than the mass of the cluster? The dense cluster environments likely disrupt/quench small galaxies and in turn strongly affect the faint end slope of the luminosity function. 

To address these questions, we started an H$\alpha$ NB observing campaign to study the large scale structure around a statistically-significant set of $19$ low-redshift ($0.15<z<0.31$) clusters sampling a range of masses, luminosities and relaxation states. In this first paper, we present the cluster sample, the survey strategy, data collection and reduction. We also discuss H$\alpha$ luminosity functions for different redshift bins, cluster merger states, masses, X-ray luminosities as well as for different environments in and around the clusters.

\renewcommand{\arraystretch}{1.2}
\begin{table*}
\begin{center}
\caption{List of targets with coordinates, redshift, X-ray luminosity, mass ($M_{200}$ estimated from weak lensing when available or total mass computed from the cluster's velocity dispersion $\sigma$) and relaxation state.}
\vspace{-5pt}
\begin{tabular}{l c c l c r c c c}
\hline\hline
Field & RA & DEC & z & $L_\mathrm{X, 0.1-2.4keV}$ & $M_{200}$ WL & $M_{total}$ $\sigma$ & State \\ 
 & $hh$ $mm$ $ss$ & $^{\circ}$ $'$ $''$ & & [$10^{44}$\,erg\,s$^{-1}$] & [$10^{14} M_{\sun}$] & [$10^{14} M_{\sun}$] & \\ \hline
A1689   &   $13^{h}11^m29^s$    &   $-01^{\circ}20'17''$    &   0.183   & 
$14$    &  $18^{+4}_{-3}$  &  $20^{+5}_{-3}$  &  relaxed \\
A963	&	$10^h17^m13^s$	&	$+39^{\circ}01'31''$	&	0.206	& \phantom{0}$6$	& $7.6^{+1.5}_{-1.3}$ &	& relaxed	\\
A1423	&	$11^h57^m17^s$	&	$+33^{\circ}36'37''$	&	0.213	& \phantom{0}$6$	& $4.6^{+1.2}_{-1.0}$	&	& relaxed	\\
A2261	&	$17^h22^m27^s$	&	$+32^{\circ}07'58''$	&	0.224	& $11$	& $12.7^{+2.3}_{-1.5}$	& & relaxed	\\
A2390	&	$21^h53^m35^s$	&	$+17^{\circ}41'12''$	&	0.228	& $13$	&	$11.1^{+1.9}_{-1.7}$ &	& relaxed, mini-halo	\\
Z2089	&	$09^h00^m36^s$	&	$+20^{\circ}53'39''$	&	0.2343	& \phantom{0}$7$	&	$\sim5$	&	& relaxed	\\
RXJ2129 &    $21^h29^m38^s$ &   $+00^{\circ}05'39''$    &   0.235   &  $12$ &
$5.3^{+1.8}_{-1.4}$ & - & relaxed, mini-halo \\
RXJ0437	&	$04^h37^m10^s$	&	$+00^{\circ}43'38''$	&	0.285	& $9$ &	 $\sim5$	&	& relaxed	\\ \hline
A545	&	$05^{h}32^m23^s$	&	$-11^{\circ}31'50''$	&	0.154	& \phantom{0}$5$	& -- & $11-18$	&	halo	\\
A3411	&	$08^{h}41^m54^s$	&	$-17^{\circ}29'05''$	&	0.169	& \phantom{0}$5$	& -- & $23-37$	&	relic	\\
A2254	&	$17^{h}17^m40^s$	&	$+19^{\circ}42'51''$	&	0.178	& \phantom{0}$5$	& --	& $15-29$	&	halo	\\
`Sausage' &	$22^{h}42^{m}50^{s}$ &	$+53^{\circ}06'30''$	&	0.188	& \phantom{0}$7$ & $25.1\pm5.3$ 	& $\sim30$	&	relic	\\
A115	& $00^h55^m59^s$	& $+26^{\circ}22'41''$	&	0.1971	& \phantom{0}$9$	& $6.7^{+3.2}_{-2.1}$	&	& relic	\\
A2163   & $16^h15^m34^s$    & $-06^{\circ}07'26''$  &   0.203   &  $38$            & $29.0^{+4.6}_{-5.8}$  & $39\pm4$ & halo \\
A773	&	$09^h17^m59^s$	&	$+51^{\circ}42'23''$	&	0.217	& \phantom{0}$6$ &	$10.2^{+1.5}_{-1.3}$	& $12-27$	& halo	\\
`Toothbrush'& $06^{h}03^{m}30^{s}$ & $+42^{\circ}17'30''$	&	0.225	& \phantom{0}$8$	& $9.6^{+2.1}_{-1.5}$ 	& $\sim{22}$ &	relic, halo	\\
A2219	&	$16^h40^m21^s$	&	$+46^{\circ}42'21''$	&	0.2256	& $12$	& $10.9^{+2.2}_{-1.8}$	&	& halo 	\\
A1300	&	$23^h23^m07^s$	&	$+01^{\circ}43'16''$	&	0.3072	& $13$	&		& $\sim6$	& halo, relic	\\
A2744	&	$00^h14^m18^s$	&	$-30^{\circ}23'22''$	&	0.308	& $13$ &	$20.6\pm4.2$	&	& halo, relic	\\
\hline
\end{tabular}
\vspace{-10pt}
\label{tab:clusters}
\end{center}
\end{table*}
\renewcommand{\arraystretch}{1.1}

The paper is organised in the following way: in Section \ref{sec:sample} we present the sample of clusters and their properties; in Section \ref{sec:obs-reduction} we discuss the NB and corresponding subtraction BB observations and their reduction, as well as any ancillary data we are using. Section \ref{sec:selection} covers the H$\alpha$ emitter selection, while in Section \ref{sec:LHA} we present the formalism of obtaining luminosity functions. In Section \ref{sec:results} we present the different H$\alpha$ luminosity functions for clusters and the fields around them binned by cluster mass, luminosity, redshift, merger stage etc. In Section \ref{sec:discussion} we discuss the implications of our results for the cosmic evolution of cluster and field galaxies. The conclusions can be found in Section \ref{sec:conclusion}. 

We assume a flat $\Lambda$CDM cosmology, with $H_{0}=70$\,km\,s$^{-1}$\,Mpc$^{-1}$, matter density $\Omega_M=0.3$ and dark energy density $\Omega_{\Lambda}=0.7$. We have made use of the online cosmology calculator presented in \citet{2006PASP..118.1711W}, as well as its \textsc{python} implementation. Images are in the J2000 coordinate system. Magnitudes are in the AB system. We use a Chabrier initial mass function \citep[IMF;][]{2003PASP..115..763C}.

\section{Cluster sample}\label{sec:sample}

\begin{figure*}
\centering
\includegraphics[trim=0cm 0cm 0cm 0cm, width=0.45\textwidth]{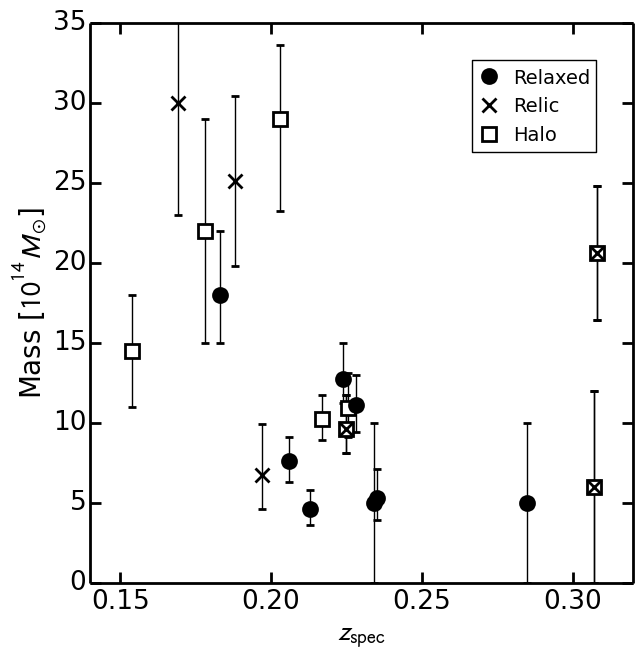}
\includegraphics[trim=0cm 0cm 0cm 0cm, width=0.45\textwidth]{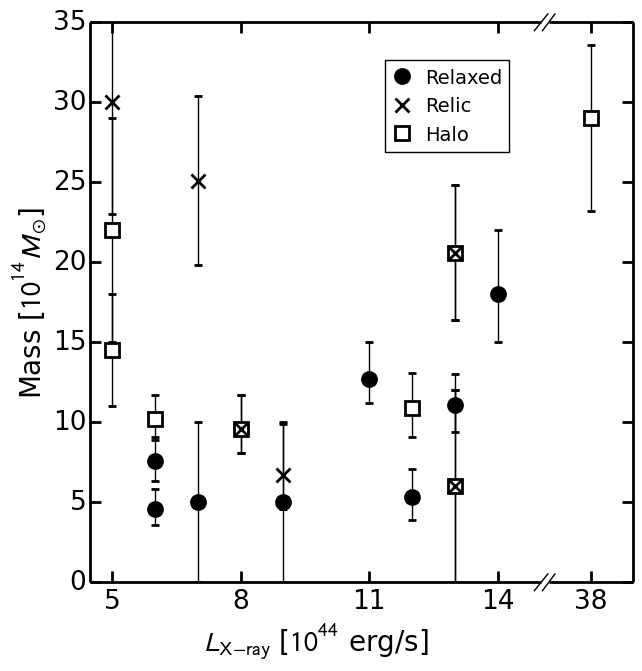}
\vspace{-10pt}
\caption{Distribution of galaxy clusters with respect to mass and redshift (left panel) and with respect to mass and X-ray luminosity (right panel). The relaxation state is encoded in the symbol. Note that masses are inferred from weak lensing estimates when available, but in some cases such an estimate was not available so we use the total mass estimate based on the cluster's velocity dispersion. Note the lack of correlation between mass and luminosity, especially for the disturbed clusters.}
\label{fig:clusters}
\end{figure*}

Our sample of $19$ clusters was selected mainly to probe a range in redshift ($0.15<z<0.31$), mass, luminosity and merger states. Our sample includes relaxed and merging clusters hosting increased turbulence and shock waves (see Figure~\ref{fig:clusters}). Increased turbulence in the ICM is indicated by the presence of diffuse radio emission co-located with the ICM \citep[halo,][]{2012A&ARv..20...54F}. ICM shocks, thought to be produced at the merger of two massive clusters, can lead to particle acceleration which in the presence of magnetic fields leads to radio synchrotron emission \citep[relics,][]{2012A&ARv..20...54F}. ICM shocks can also be detected as temperature or density discontinuities in the ICM, using X-ray data \citep[e.g.][]{2002ApJ...567L..27M}. Theory predicts that as the clusters pass through each other, the shocks are produced first, hence the relics are visible first. The merger also induces large bulk motions, which take time to cascade down to small scale ($10-100$\,kpc) turbulence capable of re-accelerating electrons and hence produce a radio halo \citep[e.g.][]{2014IJMPD..2330007B, 2013MNRAS.429.3564D}. Therefore, on average, mergers with relics only could be younger than disturbed clusters hosting a halo+relic or a halo only. Even some relaxed clusters can show some degree of disturbance at their cores: gas sloshing around the central radio galaxy in turn generates turbulence. This turbulence can re-accelerate plasma from the radio galaxy to form extended diffuse radio emission, called a mini-halo \citep{2010ApJ...717..908Z, 2012A&ARv..20...54F}.

Details about each cluster can be found in Appendix \ref{appendix}, and the main physical properties can be found summarised in Table~\ref{tab:clusters} and visualised in Figure~\ref{fig:clusters}. The targets are separated in relaxed and merging, and presented in increasing redshift order.

\begin{table*}
\begin{center}
\caption{List of targets with the luminosity distance ($D_\mathrm{L}$), NB \& BB filters used, the effective NB observing time, as well as observing period. The final column lists the volume in each field, amounting to a total volume of $1.3\times10^5$ Mpc$^3$.}
\vspace{-5pt}
\begin{tabular}{l c c c c c c}
\hline\hline
Field & $D_\mathrm{L}$ & NB filter & BB filter & NB Eff. int. time & Obs period & Volume \\ 
 & [Mpc] & (H$\alpha$) & (rest-frame R)  & [ks] & & [$10^4$\,Mpc$^3$] \\ \hline
A1689 &  887.8   &  NOVA7743 	& WFCSloanI	&   14.6     &   Jun 2016    &  \phantom{0}4.3 \\
A963 &	1013	&	NOVA7941	& WFCSloanI	&	12.6	&	Mar, Apr 2016	&	\phantom{0}5.9	\\
A1423 &	1051.6	&	NOVA7941	& WFCSloanI	&	13.8	&	Mar, Apr 2016	&	\phantom{0}6.0	\\
A2261 & 1110.5	&	NOVA804HA	& WFCSloanI	&	12	&	Jul 2015	&	\phantom{0}5.0	\\
A2390 &	1133	&	NOVA804HA	& WFCSloanI	&	15	&	Jul 2015	&	\phantom{0}4.8	\\
Z2089 &	1168.6	&	NOVA8089	& WFCSloanI	&	18	&	Oct 2012, Nov 2013 	&	\phantom{0}7.3	\\
RXJ2129 & 1103.6  & NOVA8089  & SDSS i & 7 & Jun 2016 & \phantom{0}7.1 \\
RXJ0437 &	1462.5	&	MB837 &	BBIc	&	18	&	Dec 2014	&	14.1	\\ \hline
A545 &	731.3	&	MB753 &	CFHT i	&	18	&	Dec 2014	&	\phantom{0}3.7	\\
A3411 &	808.7	&	MB770	&	Subaru i	&	18	&	Dec 2014	&	\phantom{0}4.3	\\
A2254 &	858.3	&	NOVA7743 	& WFCSloanI	&		15	&	Jul 2015	&	\phantom{0}4.5	\\
`Sausage' &	867.7	&	NOVA782HA	&	WFCSloanI	&	47.4	&	Oct 2012, Nov 2013
	&	\phantom{0}3.4	\\
A115	& 961.6 &	NOVA782HA	& SDSS i	&	16.8	&	Oct, Nov 2015	&	\phantom{0}3.6	\\
	&	& 	NOVA7941 & &	10.2	&		&	\phantom{0}5.8	\\
	&	&	NOVA7743 & 	&	11.4	&		&	\phantom{0}4.2	\\
A2163 & 996.5 & NOVA7941 &  WFCSloanI & 26.3 & Mar, Apr, Jun 2016 &  \phantom{0}6.1 \\
A773 & 1012.4	&	NOVA7941	& WFCSloanI	&	7.8	&	Nov 2015	&	\phantom{0}6.0	\\
`Toothbrush' & 1118.5	&	NOVA804HA & WFCSloanI	&		51	&	Oct 2012, Nov2013 	&	\phantom{0}4.6	\\
A2219 &	1119.6	&	NOVA804HA	& WFCSloanI	&	12	&	Jul 2015	&	\phantom{0}4.9	\\
A1300 &	1595.2	&	MB856 &	BBIc	&	18	&	Dec 2014	&	11.5	\\
A2744 &	1600	&	MB856 &	BBIc	&	18	&	Dec 2014	&	10.9	\\
\hline
\end{tabular}
\vspace{-10pt}
\label{tab:obs}
\end{center}
\end{table*}

\begin{figure*}
\centering
\begin{subfigure}[b]{0.49\textwidth}
\includegraphics[trim=0cm 0cm 0cm 0cm, height=0.90\textwidth]{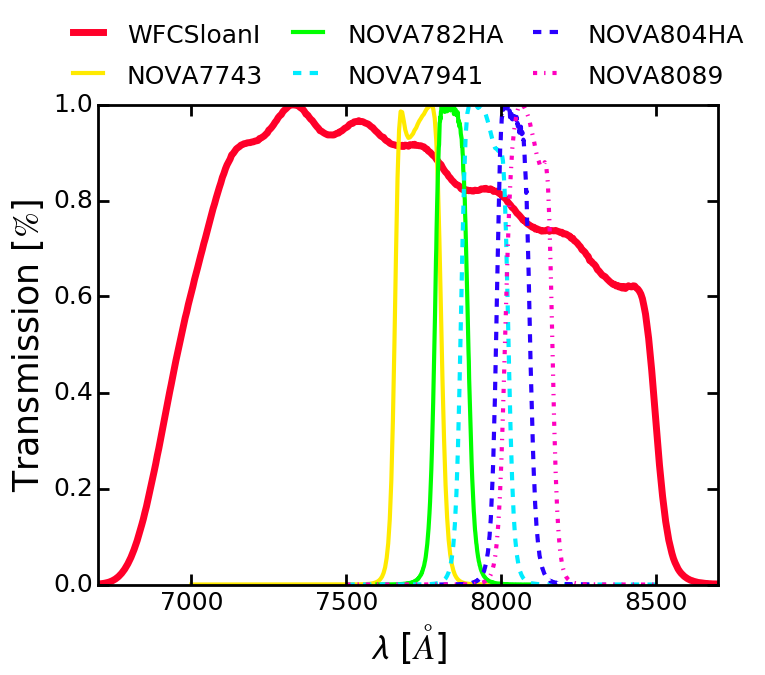}
\end{subfigure}
\hspace{5pt}
\begin{subfigure}[b]{0.49\textwidth}
\includegraphics[trim=0cm 0cm 0cm 0cm, height=0.90\textwidth ]{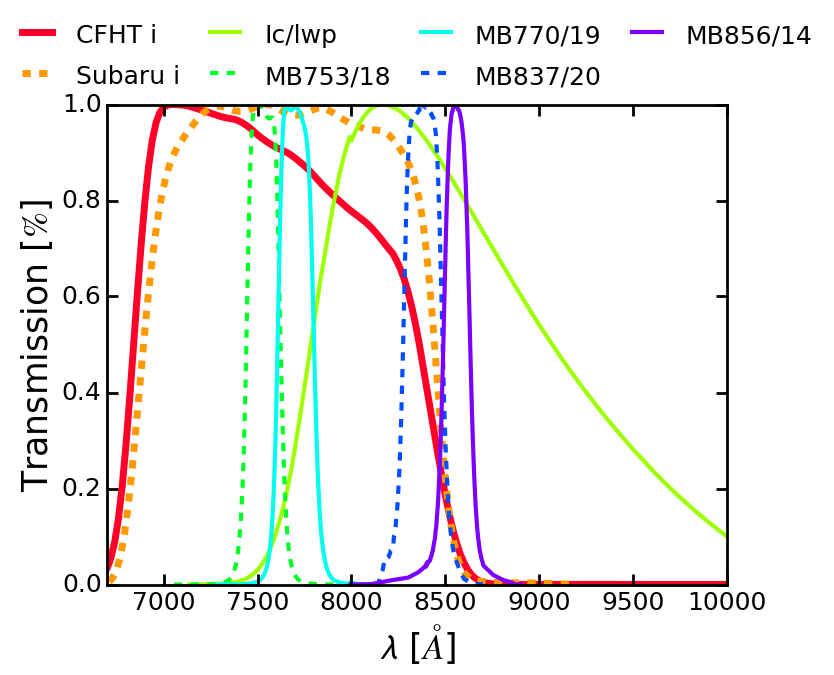}
\end{subfigure}
\vspace{-10pt}
\caption{Normalised profiles of the NB filters used to survey H$\alpha$ emitters at the redshift of our clusters. The BB filters used for continuum subtraction are also overplotted.}
\label{fig:filters}
\end{figure*}

\begin{table}
\begin{center}
\caption{Filter effective central wavelength and full-width-at-half-maximum for the filters used in this study.}
\vspace{-5pt}
\begin{tabular}{l l c c }
\hline\hline
Telescope & Filter & $\lambda_\mathrm{c}$ [{\AA}] & FWHM [{\AA}] \\ \hline
\multirow{8}{*}{INT}	&	NOVA7743	&	7731.9	&	152.5	\\
	&	NOVA782HA	&	7838.8	&	110.0	\\
	&	NOVA7941	&	7944.5	&	155.0	\\
	&	NOVA804HA	&	8037.7	&	110.5	\\
	&	NOVA8089	&	8086.7	&	152.5	\\
 	&	WFCHARB	&	4361.2	&	1020.0	\\
	&	WFCHARR	&	6505.6	&	1405.0	\\
   &	WFCSloanI	&	7671.3	&	1510.0	 \\ \hline
\multirow{5}{*}{MPG2.2}	&	MB753	&	7530.4	&	182.5	\\
	&	MB770	&	7704.1	&	192.5	\\
	&	MB837	&	8377.6	&	210.0	\\
	&	MB856	&	8557.8	&	144.0	\\
	&	BBIc	&	8299.7	&	1283.8	\\ \hline
\multirow{5}{*}{CFHT} &	u	&	3798.7	&	700.0	\\
	&	g	&	4861.0	&	1430.0	\\
	&	e	&	6260.1	&	1220.0	\\
	&	I	&	7577.4	&	1520.0	\\
	&	z	&	8876.2	&	870.0	\\ \hline
\multirow{3}{*}{Subaru}	&	g	&	4794.2	&	1174.3	\\
	&	r	&	6263.2	&	1414.4	\\
	&	I	&	7666.5	&	1542.5	\\
\multirow{4}{*}{SDSS}	&	g	&	4640.4	&	1158.4	\\
	&	r	&	6122.3	&	1111.2	\\
	&	I	&	7439.5	&	1044.6	\\
	&	z	&	8897.1	&	1124.6	\\
\hline
\end{tabular}
\vspace{-10pt}
\label{tab:filters}
\end{center}
\end{table}

\section{Data, Observations \& Data Reduction}
\label{sec:obs-reduction}

\subsection{Ancillary data}\label{sec:ancillary}

Our targets have useful ancillary data in the form of additional targeted or public survey photometry or spectroscopic redshifts. Note however that the photometry and spectroscopy availability and quality is highly dependent on the field, thus resulting in inhomogeneous ancillary data.

Many of the clusters are covered by the SDSS survey in its 9th data release \citep[SDSS DR9;][]{2009ApJS..182..543A}. For A2744, we employ the VLT Survey Telescope ATLAS survey data available in the g, r, i and z bands \citep{2015MNRAS.451.4238S}. Four clusters have fully reduced and stacked images produced using the MegaPipe image stacking pipeline which are made available through Terapix\footnote{A545: g, r, i, z bands, PI Morrison, ID 05BH42; A1300: g, r bands, PI Richard, ID 13AF05; A2163: g, r bands, PI Hoekstra, 05AC10; RXJ2129: g, r, i bands, PI Kneib, 10BF23 and PI Rogerson, 12BC31}. We also employ g, r, i Subaru images of A3411 presented in \citet{weeren2016}. We downloaded BB data available from the INT and ESO/MPG 2.2m archives and reduced in the manner described below in Section~\ref{sec:red}. For A115 and RXJ2129, we used the SDSS i band data for BB subtraction mosaicked through {\sc Montage}\footnote{\url{http://hachi.ipac.caltech.edu:8080/montage/index.html}}, which we processed in the same way as all of the other data (see Section~\ref{sec:red}). 

For near infra-red (NIR) bands, we make use of data from the Visible and Infrared Survey Telescope for Astronomy (VISTA) Hemisphere Survey \citep[VHS;][]{2013Msngr.154...32E} and the VISTA Kilo-degree Infrared Galaxy Survey \citep[VIKING;][]{2013Msngr.154...35M}, as well as the United Kingdom Infra-Red Telescope (UKIRT) Infrared Deep Sky Survey data \citep[UKIDSS;][]{2007MNRAS.379.1599L}. When such deep data are not available, we explore all sky NIR data from the 2MASS survey \citep{2006AJ....131.1163S}.

We collect redshifts available from targeted studies on particular clusters in our samples \citep{1997A&A...326...34L, 1997A&AS..124..283P, 2004A&A...416..839B, 2004A&A...425..797L, 2007A&A...467...37B, 2007A&A...469..861B, 2007ApJ...665..921F, 2008A&A...481..593M, 2011A&A...529A.128B, 2011A&A...536A..89G, 2011ApJ...728...27O, 2012ApJ...757...22C, 2012MNRAS.423..256H, 2012MNRAS.420.2480Z, 2013ApJ...776...91L, 2015ApJ...805..143D, 2015MNRAS.450..630S, 2016ApJ...817..179J, weeren2016}. We also make use of the redshift compilation from \citet{2013ApJ...767...15R} and the 2dF Galaxy Redshift Survey \citep[2dFGRS;][]{2001MNRAS.328.1039C}. Note however that most of these studies specifically targeted the passive galaxy population, thus we do not necessarily expect overlap with the sources we will select as H$\alpha$ emitters. Additionally, we do not have many redshifts for sources at other than the cluster redshift. However, these data are useful to check the reliability of our star-forming galaxy selection methods (i.e. galaxies confirmed as passive with spectroscopy should not be selected as H$\alpha$ emitters). The spectroscopic redshifts are used in Section~\ref{sec:selection}.

\subsection{New H$\alpha$ NB and associated BB observations}\label{sec:obs}

We acquired NB data tracing H$\alpha$ emission in the field and at the redshift of each cluster, as well as associated BB observations. The survey is designed to capture a sufficiently large field of view (FOV, $\sim0.5$ deg$^2$) in a single exposure to avoid inhomogeneities caused by mosaicking. At full depth, the survey reaches galaxies a few orders of magnitude fainter than typical H$\alpha$ emitters, whilst still capturing the brightest H$\alpha$ emitters. We targeted clusters to match existing NB filters mounted on wide-field cameras. Additionally, we built custom made NB filters to cover specific redshift slices, optimised to capture H$\alpha$ emission at the redshift of a few clusters. We compare the redshift range covered by the clusters given their velocity dispersion $\sigma$ and find all clusters but A2163 are fully covered within 1.644$\sigma$ from the central redshift. Within this $1.664\sigma$ range, we encompass 90 per cent of cluster galaxies and the cut will happen only at one side of the distribution. Therefore, for all clusters but A2163 we cover at least 95 per cent of the cluster line emitters. Because of its high mass and large velocity dispersion, the lower redshift distribution of A2163 galaxies is not fully covered by the NB filter. The filter covers down to $-1\sigma$. This amounts to covering at least 85 per cent of cluster sources. Therefore, as per our design, the filters cover very well the redshift distribution of clusters.

\subsubsection{Isaac Newton Telescope data}
For the northern targets, we used the Wide Field Camera (WFC)\footnote{\url{http://www.ing.iac.es/engineering/detectors/ultra_wfc.htm}} 
mounted on the 2.5-m Isaac Newton Telescope\footnote{\url{http://www.ing.iac.es/Astronomy/telescopes/int/}}. The WFC consists of four CCDs (pixel scale of 0.333\,pixel\,arcsec$^{-1}$) forming a $0.56\times0.56$\,deg$^2$ with the top-right (NW on the sky) corner missing, with chip gaps of $\sim20$\,arcsec. The observations were taken in a five-point dither pattern to cover the chip gaps. 

Data were taken over a total of 16 nights, between Jul 2015 and June 2016, with a variety of moon phases (8 dark, 3 gray and 5 bright nights) and observing conditions (seeing of $0.8"-2.0"$). We took $600$\,s individual exposures in the NB filters and $200$\,s exposures on the BB filters, to avoid saturation of bright objects. This strategy enables us to identify bright emitters as well as avoid sky area loss because of saturation halos and spikes around bright stars. To this, we are also adding data on the `Sausage' and `Toothbrush' clusters presented in \citet{2014MNRAS.438.1377S} and \citet{Stroe2015}. For many clusters, the observations were taken months apart which allows the removal of variable and moving sources through stacking. 

For each cluster, we obtained data in one NB filter chosen to cover the H$\alpha$ emission redshifted at the distance of each galaxy cluster. The only exception is A115, where we took NB observation in 3 NB redshift slices to cover the H$\alpha$ emission in sources in the foreground, inside and in the background of the cluster. We used the already existing custom-made NB filters presented in \citet{2014MNRAS.438.1377S}, NOVA782HA and NOVA804HA. We also bought new custom-made filters (NOVA7743, NOVA7941, NOVA8089) of about $~150$\,{\AA} width. A total of 5 separate NB filters were used for this study. With our 5 filters, we have continuous H$\alpha$ coverage between $z\sim0.166$ and $z\sim0.244$. 

The details of the NB filters and other BB filter data we employed can be found in Table~\ref{tab:filters} and Figure~\ref{fig:filters}. The filter profiles have been convolved with the quantum efficiency of the CCD and the effect of the optics. In case of the clusters observed with the INT, we obtained data in the WFCSloanI filter to measure the continuum emission. For A115, we used SDSS images to extract sources for BB subtraction in the same way as all the other images. The exact filters used as NB and BB for broad emission subtraction for each cluster are listed in Table~\ref{tab:obs}.

\subsubsection{ESO2.2m telescope data}
For the southern targets, we used the Wide Field Imager \citep[WFI]{1999Msngr..95...15B} on the ESO/MPG 2.2-m telescope\footnote{\url{http://www.mpia.de/science/2dot2m}}. Eight individual $2k\times4k$ CCDs (with $0.238$\,arcsec pixel scale) form the detector, with $14$\,arcsec and $23$\,arcsec chip gaps in the NS and EW directions, respectively. A seven-point dither pattern was employed obtain contiguous sky coverage across the chip gaps. 

The data were taken in excellent seeing conditions ($0.4"-0.6"$) in Dec 2014, under dark skies using 4 different NB filters to match the redshifts of the clusters. With the NB filters we cover the $0.133-0.189$ redshift range and the $0.260-0.315$ range. As with the INT data, NB filter exposures were $600$\,s, with $200$\,s for the BB. Observations in the filter BBIc were taken for BB subtraction. However, in the case of some clusters, this filter is too red, so CFHT (available from Terapix) and Subaru \citep{weeren2016} i-band images were used. Table~\ref{tab:obs} lists the details of which NB and which BB filter was used for each cluster.

\begin{table}
\begin{center}
\caption{Clusters with NB and BB filters. Average $3\sigma$ limiting magnitudes (measured in $5$\,arcsec apertures) for the different fields in the NB and BB. The limits are calculated per chip and we report the average. We also add a standard deviation of these limits, which is calculated between the values obtained for the different chips. We also report the limiting H$\alpha$ luminosity at 50 per cent completeness, as well as the total number of emitters selected in each field.}
\vspace{-5pt}
\begin{tabular}{l l c c c c}
\hline\hline
Field & Filter & Avg.   & Std. dev.  & Lim.  & No.  \\ 
      &        & $3\sigma$ & $3\sigma$ & $\log L$ & emit. \\
& & [mag] & [mag] & [erg s$^{-1}$] &  \\ \hline
\multirow{2}{*}{A1689}	&  NOVA7743 & 20.16	& 0.05 &  40.2 & 291 \\
& WFCSloanI & 19.81 & 	0.07 \\\hline
\multirow{2}{*}{A963}	&	NOVA7941	&	20.16	&	0.03 & 40.5 & 100	\\ 
&	WFCSloanI	&	21.23	&	0.08 &  &	\\ 	 \hline
\multirow{2}{*}{A1423}		&	NOVA7941	&	20.12	&	0.07& 40.4 & 193\\ 
	&	WFCSloanI	&	21.12	&	0.08 &  & \\  \hline
\multirow{2}{*}{A2261}		&	NOVA804HA	&	19.95	&	0.03 & 40.5	& 361 \\
&	WFCSloanI	&	19.89	&	0.04 & &	\\  \hline
\multirow{2}{*}{A2390} 	&	NOVA804HA	&	20.2	&	0.04 & 40.4 & 258	\\ 
		&	WFCSloanI	&	20.52	&	0.07 & & \\  \hline
\multirow{2}{*}{Z2089}	&	NOVA8089	&	18.83	&	0.03 & 41.2 & \phantom{0}67	\\ 
		&	WFCSloanI	&	20.39	&	0.10 & & 	\\ \hline
\multirow{2}{*}{RXJ2129} & 	NOVA8089 &	19.64 &	0.03 & 41.0 & 130  \\
&	i	& 19.73 &	- & &  \\ \hline
\multirow{2}{*}{RXJ0437}	&	MB837	&	19.53	&	0.04 & 41.0 & 293 \\ 
	&	BBIc	&	19.35	&	0.0 5& & \\ 	 \hline
\multirow{2}{*}{A545}	&	MB753	&	19.75	&	0.07 & 40.4 & 148	\\ 
	&	i	&	20.66	&	-	& & \\  \hline
\multirow{2}{*}{A3411}	&	MB770	&	19.93	&	0.05 & 40.5 & 410	\\ 
	&	i	&	21.81	&	-	& & \\ \hline
\multirow{2}{*}{A2254}	&	NOVA7743	&	20.59	&	0.05 & 40.2 &  391	\\ 
&	WFCSloanI	&	21.31	&	0.02 & & 	\\ \hline
\multirow{2}{*}{`Sausage'}	&	NOVA782HA	&	18.94	&	0.20 & 40.7 & 201	\\ 
	&	WFCSloanI	&	19.08	&	0.19 & & 	\\ \hline
\multirow{4}{*}{A115}	&	NOVA782HA	&	19.57	&	0.07	& 40.6 & 144 \\ 
	&	NOVA7941	&	18.86	&	0.05 &	41.0  & \phantom{0}56 \\ 
	&	NOVA7743	&	19.08	&	0.06 & 	41.0 & \phantom{0}68 \\ 
	&	i	&	19.71	&	- & & 	\\ \hline
\multirow{2}{*}{A2163} & NOVA7941 & 19.24 & 0.08 & 40.7 & 146\\
& WFCSloanI &	20.12 &	0.04 & &  \\ \hline
\multirow{2}{*}{A773}	&	NOVA7941	&	19.17	&	0.05 & 41.0 & 140\\ 
	&	WFCSloanI	&	19.67	&	0.05 & 	& \\  \hline
\multirow{2}{*}{`Toothbrush'}	&	NOVA804HA	&	20.03	& 0.08 & 40.4 & 463 	\\ 
	&	WFCSloanI	&	20.58	&	0.06 & & 	\\  \hline
\multirow{2}{*}{A2219}	&	NOVA804HA	&	20.17	&	0.05 & 40.4 & 536	\\ 
		&	WFCSloanI	&	20.60	&	0.06 & & 	\\  \hline
\multirow{2}{*}{A1300}		&	MB856	&	19.52	&	0.08 & 40.9 & 890 	\\ 
&	BBIc	&	19.20	&	0.35 & &  \\ \hline
\multirow{2}{*}{A2744}		&	MB856	&	19.52	&	0.09 & 40.7	&  619 \\ 
&	BBIc	&	19.48	&	0.07 & & \\ 
\hline
\end{tabular}
\vspace{-10pt}
\label{tab:limmag}
\end{center}
\end{table}

\begin{figure*}
\centering
\begin{subfigure}[b]{0.33\textwidth}
\includegraphics[trim=0cm 0cm 0cm 0cm, width=0.975\textwidth]{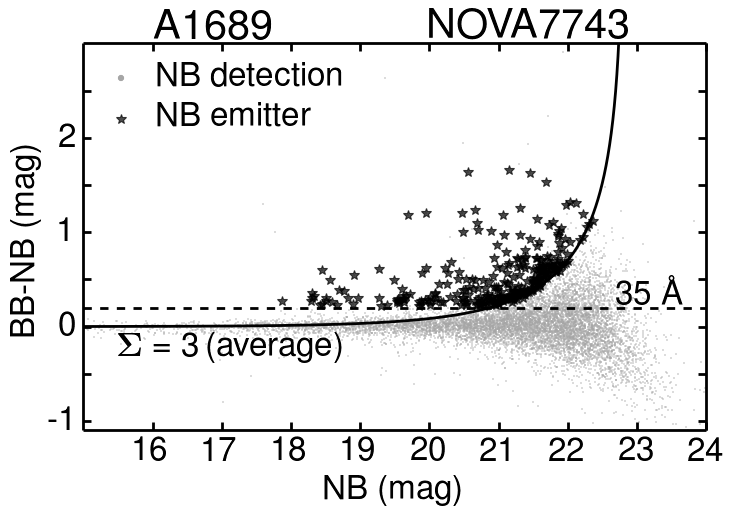}
\end{subfigure}
\begin{subfigure}[b]{0.33\textwidth}
\includegraphics[trim=0cm 0cm 0cm 0cm, width=0.975\textwidth]{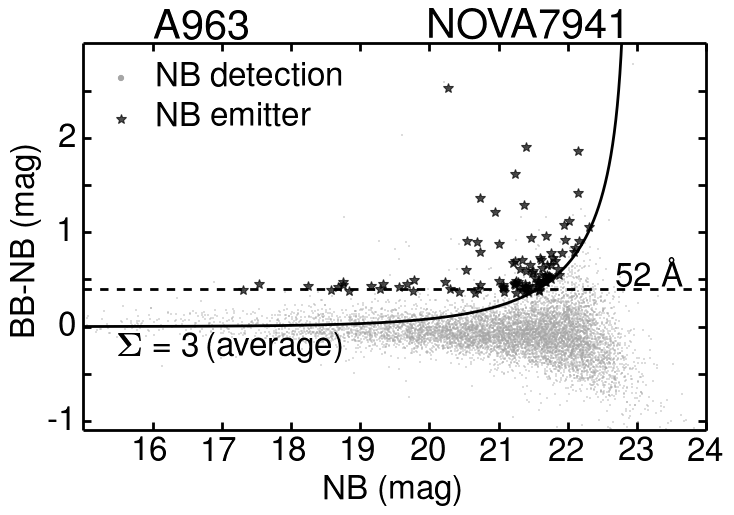}
\end{subfigure}
\begin{subfigure}[b]{0.33\textwidth}
\includegraphics[trim=0cm 0cm 0cm 0cm, width=0.975\textwidth]{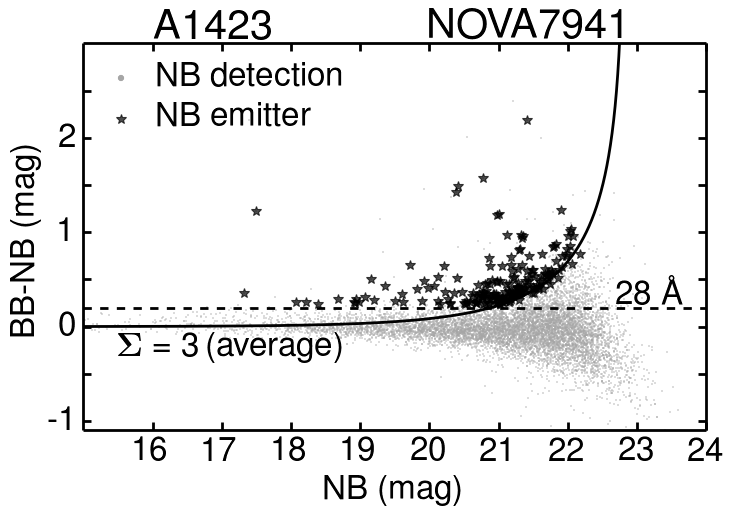}
\end{subfigure}
\\
\begin{subfigure}[b]{0.33\textwidth}
\includegraphics[trim=0cm 0cm 0cm 0cm, width=0.975\textwidth]{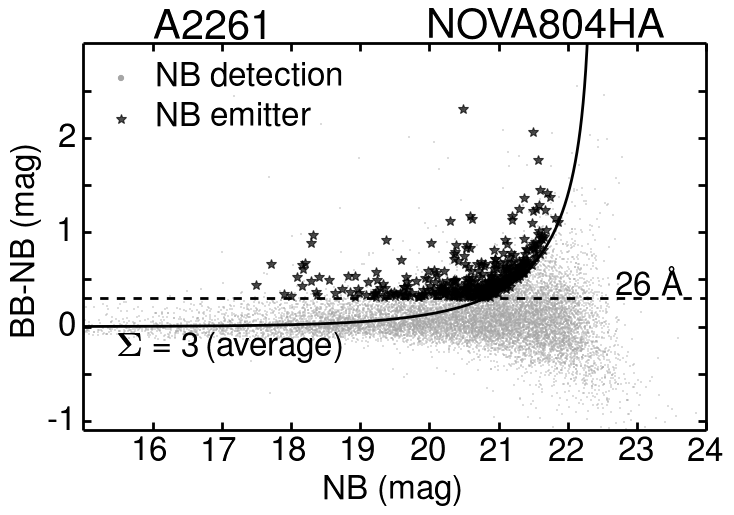}
\end{subfigure}
\begin{subfigure}[b]{0.33\textwidth}
\includegraphics[trim=0cm 0cm 0cm 0cm, width=0.975\textwidth]{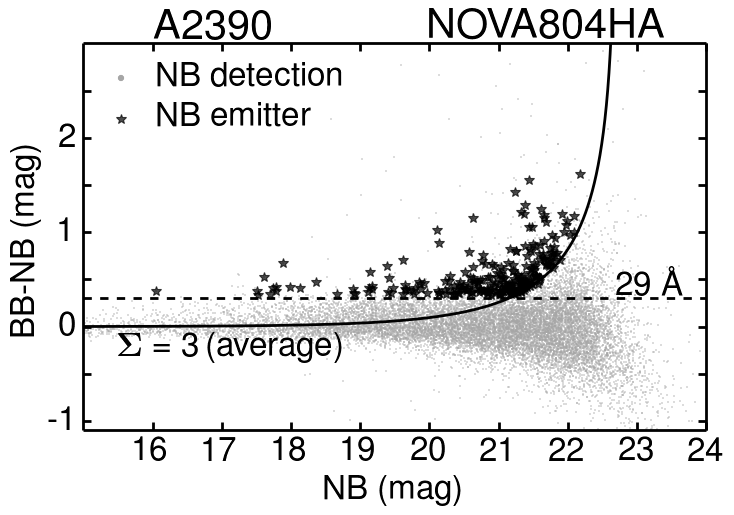}
\end{subfigure}
\begin{subfigure}[b]{0.33\textwidth}
\includegraphics[trim=0cm 0cm 0cm 0cm, width=0.975\textwidth]{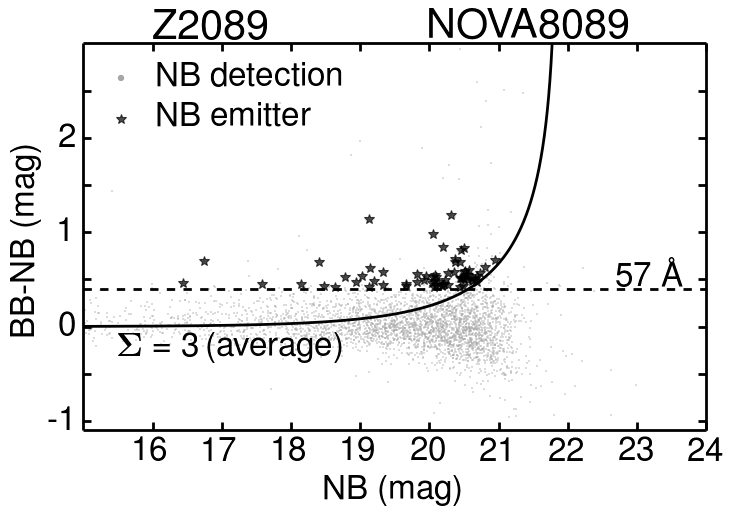}
\end{subfigure}
\\
\begin{subfigure}[b]{0.33\textwidth}
\includegraphics[trim=0cm 0cm 0cm 0cm, width=0.975\textwidth]{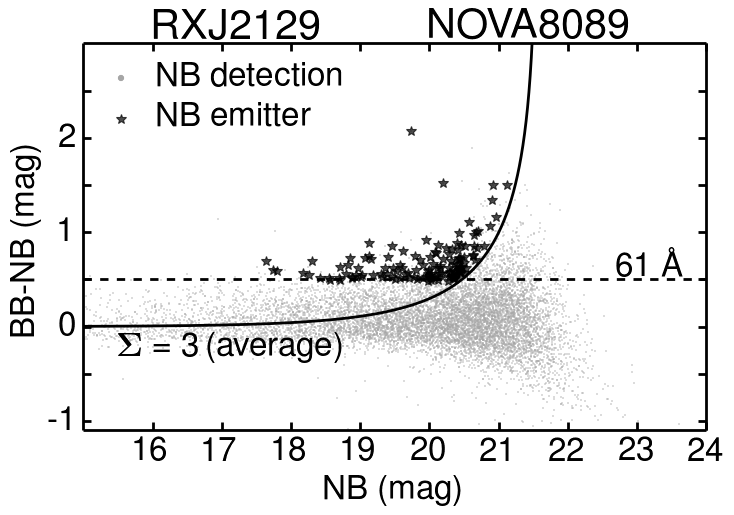}
\end{subfigure}
\begin{subfigure}[b]{0.33\textwidth}
\includegraphics[trim=0cm 0cm 0cm 0cm, width=0.975\textwidth]{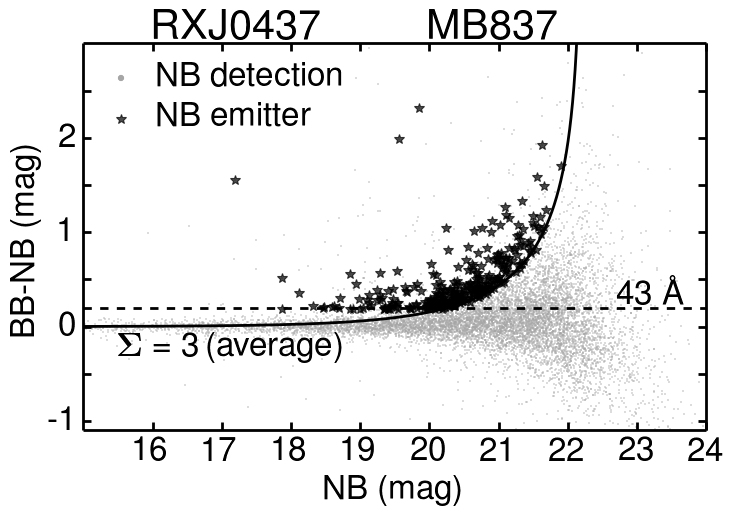}
\end{subfigure}
\begin{subfigure}[b]{0.33\textwidth}
\includegraphics[trim=0cm 0cm 0cm 0cm, width=0.975\textwidth]{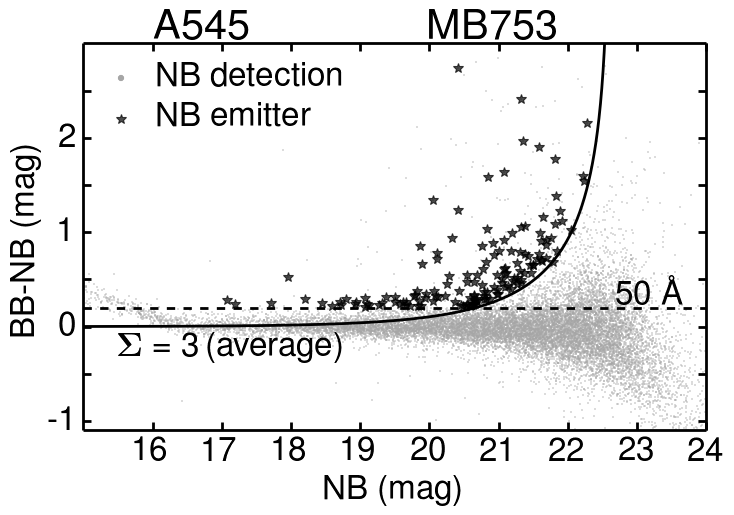}
\end{subfigure}
\vspace{-10pt}
\caption{Colour-magnitude diagrams of the colour excess as function of NB magnitude. We select emitters separately on each CCD for each cluster and adapt the cuts to reflect the noise levels reached in each observation. The curve represent the average $3\Sigma$ colour significances and the dashed, black line represents the restframe $EW_0$ cut.}
\label{fig:colmag}
\end{figure*}

\begin{figure*}\ContinuedFloat
\centering
\begin{subfigure}[b]{0.33\textwidth}
\includegraphics[trim=0cm 0cm 0cm 0cm, width=0.975\textwidth]{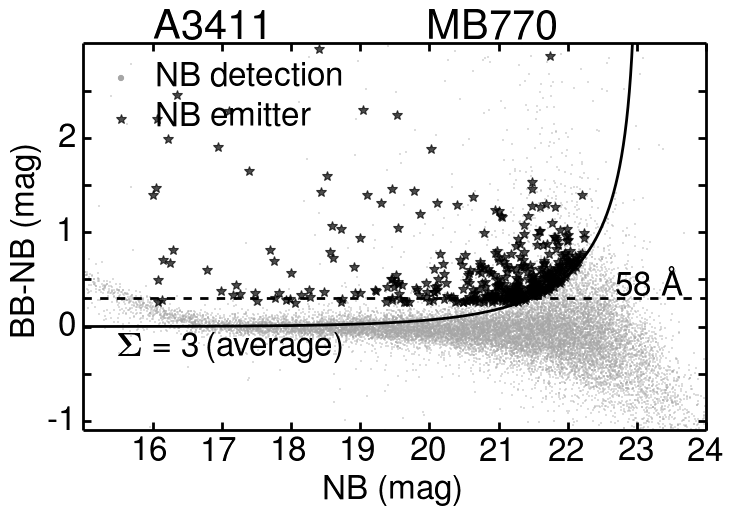}
\end{subfigure}
\begin{subfigure}[b]{0.33\textwidth}
\includegraphics[trim=0cm 0cm 0cm 0cm, width=0.975\textwidth]{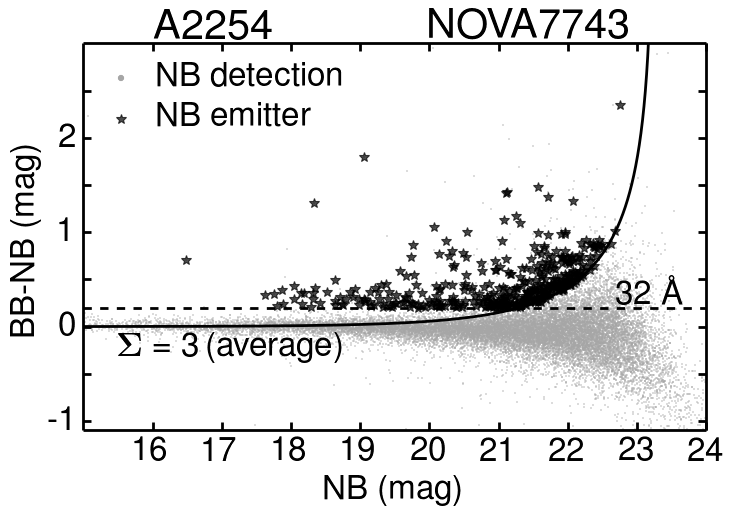}
\end{subfigure}
\begin{subfigure}[b]{0.33\textwidth}
\includegraphics[trim=0cm 0cm 0cm 0cm, width=0.975\textwidth]{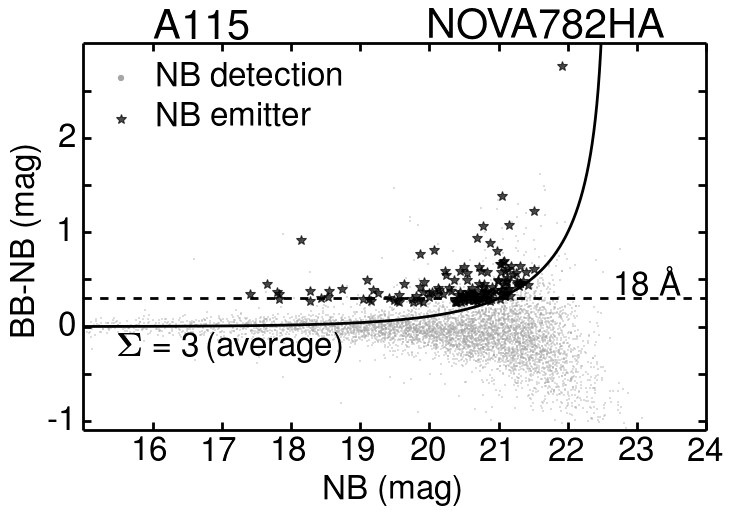}
\end{subfigure}
\\
\begin{subfigure}[b]{0.33\textwidth}
\includegraphics[trim=0cm 0cm 0cm 0cm, width=0.975\textwidth]{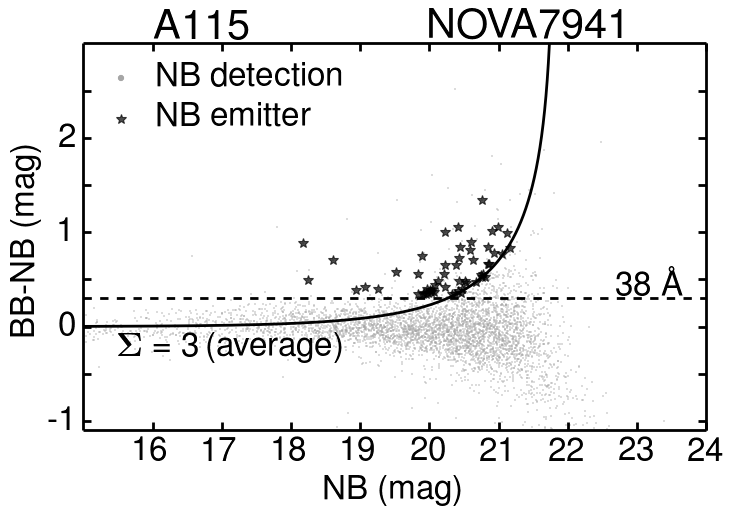}
\end{subfigure}
\begin{subfigure}[b]{0.33\textwidth}
\includegraphics[trim=0cm 0cm 0cm 0cm, width=0.975\textwidth]{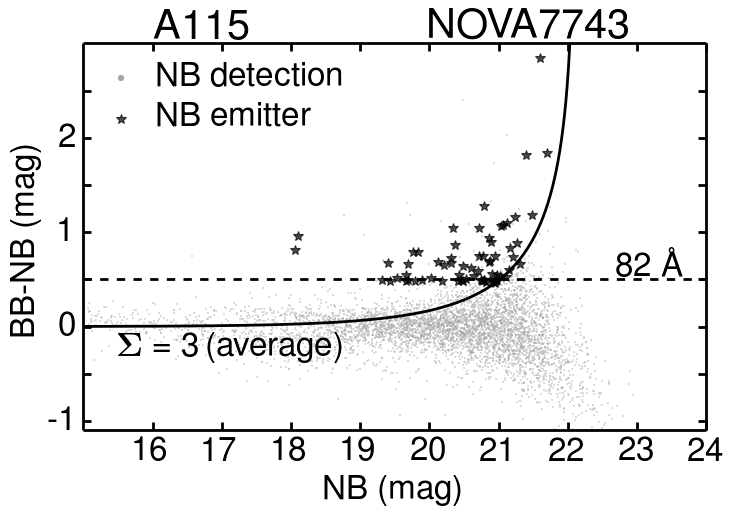}
\end{subfigure}
\begin{subfigure}[b]{0.33\textwidth}
\includegraphics[trim=0cm 0cm 0cm 0cm, width=0.975\textwidth]{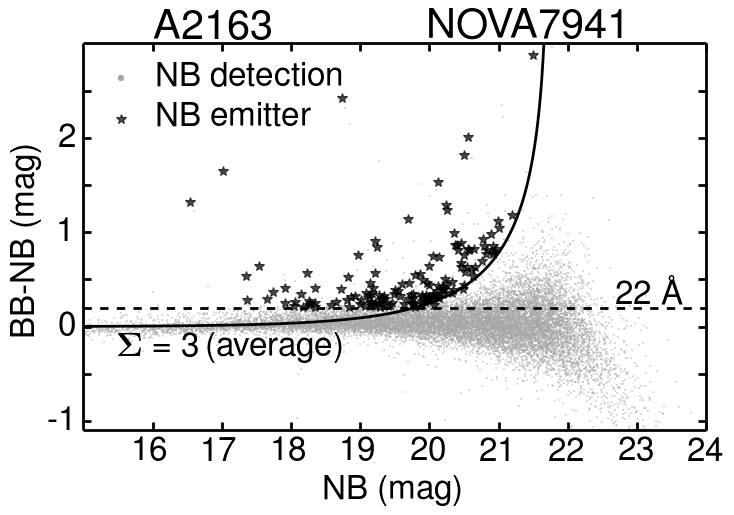}
\end{subfigure}
\\
\begin{subfigure}[b]{0.33\textwidth}
\includegraphics[trim=0cm 0cm 0cm 0cm, width=0.975\textwidth]{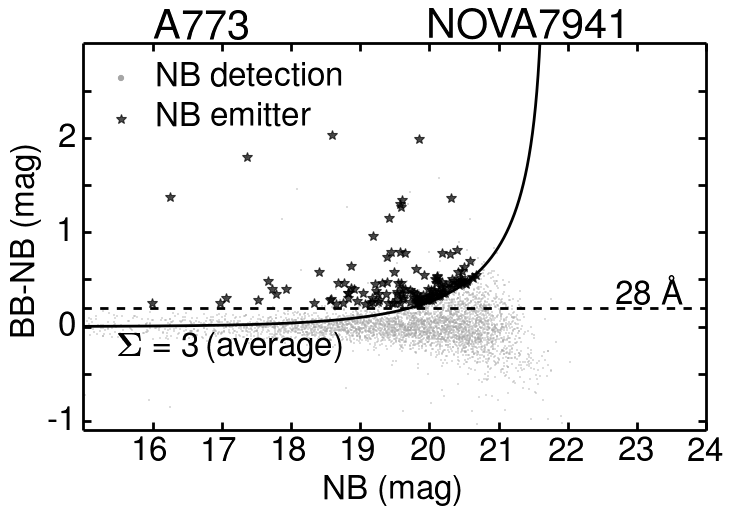}
\end{subfigure}
\begin{subfigure}[b]{0.33\textwidth}
\includegraphics[trim=0cm 0cm 0cm 0cm, width=0.975\textwidth]{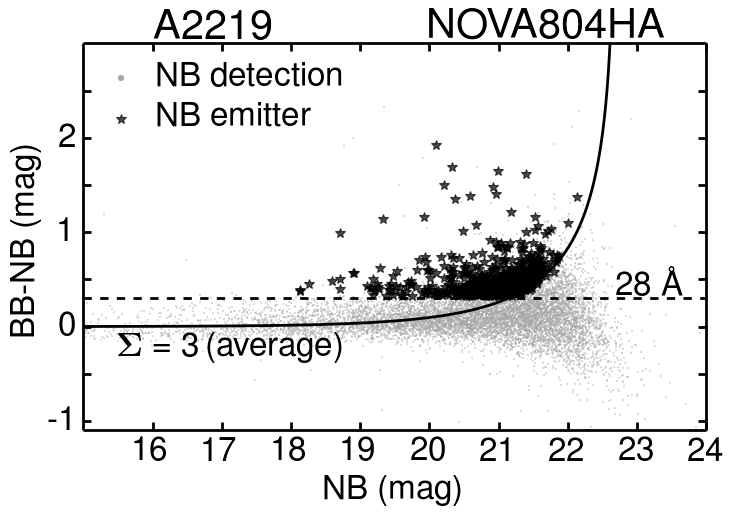}
\end{subfigure}
\begin{subfigure}[b]{0.33\textwidth}
\includegraphics[trim=0cm 0cm 0cm 0cm, width=0.975\textwidth]{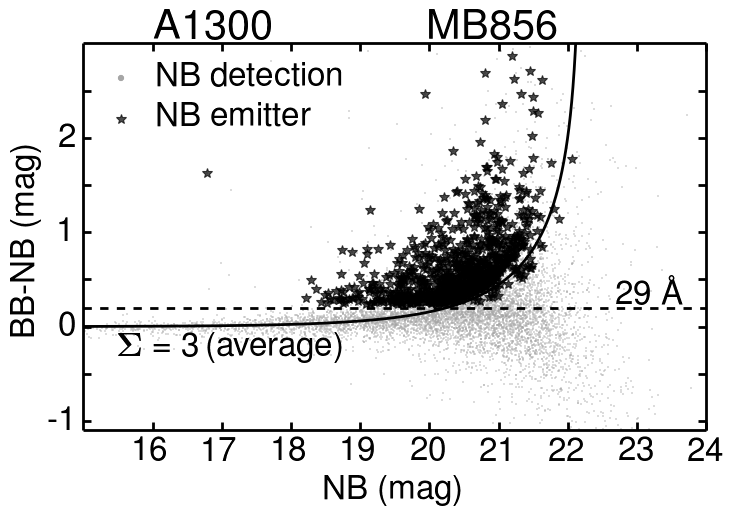}
\end{subfigure}
\\
\begin{subfigure}[b]{0.33\textwidth}
\includegraphics[trim=0cm 0cm 0cm 0cm, width=0.975\textwidth]{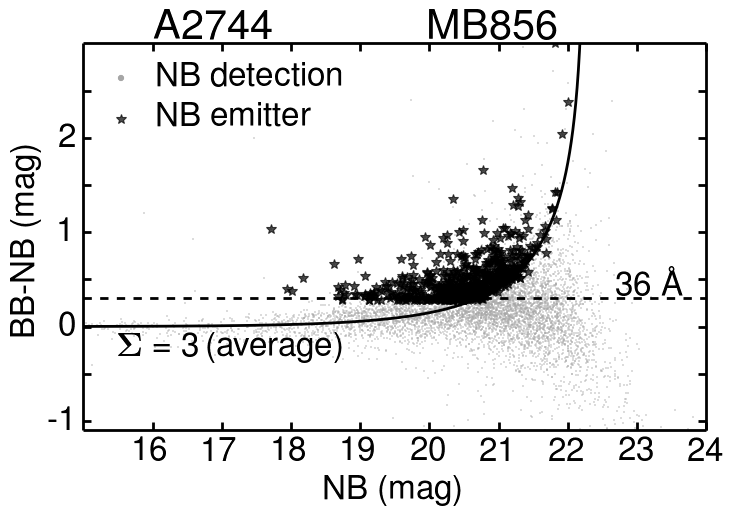}
\end{subfigure}
\vspace{-10pt}
\caption{ (cont.) Colour-magnitude diagrams of the colour excess as function of NB magnitude. We select emitters separately on each CCD for each cluster and adapt the cuts to reflect the noise levels reached in each observation. The curve represent the average $3\Sigma$ colour significances and the dashed, black line represents the restframe $EW_0$ cut.}
\end{figure*}

\subsection{H$\alpha$ NB and associated BB data reduction}\label{sec:red}

We reduced the NB and BB data using our data reduction pipeline implemented in \textsc{PYTHON} \citep{2014MNRAS.438.1377S}, in combination with the \textsc{AstrOmatic}\footnote{\url{www.astromatic.net}} software package, specifically {\sc SExtractor} \citep{1996A&AS..117..393B}, {\sc SCAMP} \citep{2006ASPC..351..112B}, {\sc SWarp} \citep{2002ASPC..281..228B} and {\sc MissFITS} \citep{2008ASPC..394..619M}.

We remove bad frames that are affected by bad weather (bad seeing, clouds, Saharan dust) and technical issues (loss of guiding, read-out issues). We also removed twilight flats which had too low or too high counts, thus being outside of the linearity range for the cameras. We median combine biases for each night to obtain a `master' bias. We subtract the overscan from the science and twilight flat frames using the `master' bias. We obtain a `master' flat by median combining the twilight flats for each filter and renormalising to 1. We correct the science frames by dividing through the `master' flat. 

In the red filters, our data suffer from `fringing', thin film interference in the CCD coating. To correct for this, we detect sources in science exposures using {\sc SExtractor} and subsequently mask them. We then median combine the masked science frames to obtain a `super-flat'. We divide the data by the `super-flat' to correct for `fringing'. 

Starting from an initial approximate astrometric solution, we use a few iterations of {\sc SCAMP} to refine the solutions over the large FOVs of our cameras. Source positions were compared with positions in the bluest band of the Two Micron All Sky Survey \citep[2MASS;][]{2006AJ....131.1163S}. {\sc MissFits} was used to update the header with the new astrometry in between {\sc SCAMP} runs. 

To bring the science exposures to the same scale, we derive zero-points (ZP) by comparing magnitudes of non-saturated objects with the closest band from the fourth United States Naval Observatory (USNO) CCD Astrograph
Catalog \citep[UCAC4;][]{2013AJ....145...44Z}. The science frames with the same ZP are median combined and background subtracted to produce final images using {\sc SWarp}.

We photometrically calibrate our data using the closest reference band in the Sloan Digital Sky Survey (SDSS) Data Release 9 \citep[SDSS DR9;][]{2012ApJS..203...21A}, when available. Some of the cluster fields are not covered by SDSS, so we use the all-sky USNO-B1.0 catalog \citep{2003AJ....125..984M}. We follow the methods described in \citet{2014MNRAS.438.1377S} to calibrate USNO-B1.0 magnitudes against the SDSS DR9 scale. We then transfer the SDSS scale to our data, using the USNO-B1.0 magnitudes as reference. We perform the photometric ZP determination for each CCD separately.

We mask saturated sources and extract magnitudes in apertures of $5$\,arcsec in diameter using {\sc SExtractor} in each CCD separately. This diameter was chosen to be large enough ($\sim15$\,kpc) to encompass the bulk of the H$\alpha$ emission at the redshifts ($0.15<z<0.31$) of our clusters. We correct all the magnitudes for Galactic dust extinction following the method described in \citet{2014MNRAS.438.1377S}, using the \citet{2011ApJ...737..103S} extinction values and interpolating to the effective wavelengths of our filters by using their model.

The average $3\sigma$ limiting magnitudes as well as the spread in the values between the different camera chips are reported in Table~\ref{tab:limmag}. The values presented are calculated after correcting for Galactic dust extinction, hence represent intrinsic depth values. Differences between the depth in each chip of the same camera are caused by variations in sensitivity and quality of the CCDs as well as the amount of Milky Way dust extinction.

\section{Selecting H$\alpha$ emitting sources}\label{sec:selection}
We cross-match the BB subtraction filter data with the NB data. We combine this catalogue with the ancillary optical, IR and spectroscopic data in order to discriminate between different types of sources and to study them in greater detail. 

\subsection{Selection of NB excess sources}\label{sec:excess}
To identify emission line systems, we first need to select sources with excess emission in NB filter compared to the BB -- this indicates the likely presence of an emission line located within the NB filter. We only select sources with a significant S/N (higher than 5). In practice, we apply these criteria using the formalism developed by \citet{1995MNRAS.273..513B}, using a colour excess significance ($\Sigma$) and an equivalent width ($EW$) cut. The colour excess significance cut ensures we select only sources with real NB excess (compared to a random scatter of colour excess), while the $EW$ cut ensures we select sources with line excess emission higher than the scatter of the excess at bright magnitudes.

Slight mismatches between the effective central wavelength of the NB filter compared to the BB can cause a systematic colour offset between magnitudes measured in the two filters. Therefore, we first correct for this effect by correcting for the median colour of sources with bright, non-saturated magnitudes. Figure \ref{fig:colmag} shows the dependence of the excess BB-NB colour on the NB magnitude, together with the $EW$ and $\Sigma$ cuts used to select emitters.
 
$\Sigma$ is then defined as \citep{2013MNRAS.428.1128S}:
\begin{equation}
\label{eq:Sigma}
\Sigma = \frac{10^{-0.4\left(m_{BB}-m_{NB}\right)}}{10^{-0.4(ZP_{AB}-m_{NB})} \sqrt{\pi r^2 \left(\sigma^2_\mathrm{NB}+\sigma^2_\mathrm{BB}\right)}},
\end{equation}
where $m_{NB}$ and $m_{BB}$ are the NB and BB magnitudes, respectively, $ZP_{AB}$ is the magnitude system zero-point, $r$ is the radius of the aperture used to extract the magnitudes measured in pixels (equivalent to $5$\,arcsec in our case) and $\sigma_\mathrm{NB}$ and $\sigma_\mathrm{BB}$ are the rms noise levels in counts, as discussed towards the end of Section \ref{sec:red}. 

The flux density $f$ is defined as: 
\begin{equation}
\label{eq:flux_broad}
f_{NB,BB} = \frac{c}{\lambda^2_{NB,BB}} 10^{-0.4(m_{NB,BB}-ZP_{AB})},
\end{equation}
where $\lambda$ is the effective central wavelength of the NB and BB filters, respectively, and $c$ is the speed of light. 
\begin{figure}
\begin{center}
\includegraphics[trim=0cm 0cm 0cm 0cm, width=0.495\textwidth]{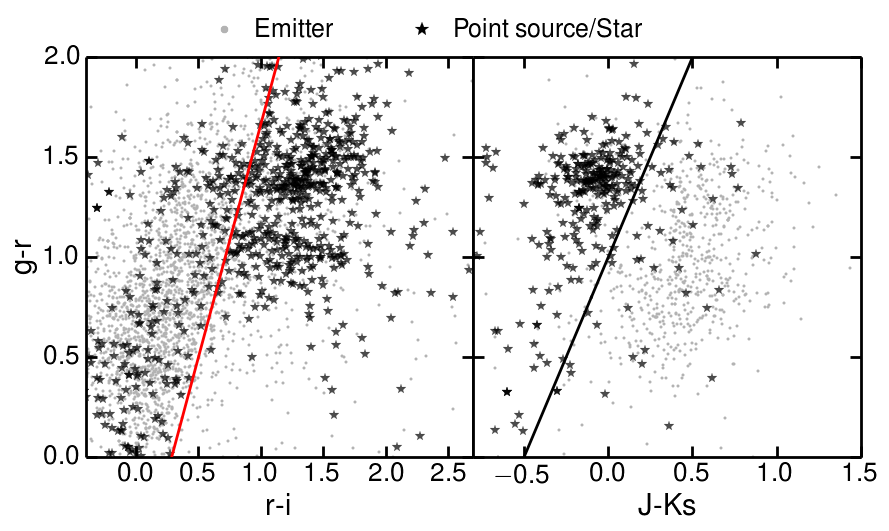}
\end{center}
\vspace{-10pt}
\caption{Colour-colour plots for emitters. The left plot displays the $g-r$ colour versus $r-i$. The right plot represents $g-r$ versus $J-Ks$. Point-like objects are represented with stars, while emitters are shown in gray dots. The lines show the colour cuts used to select point-like objects, in combination with other criteria as discussed in Section \ref{sec:stars}.}
\vspace{-10pt}
\label{fig:colcol}
\end{figure}
The line flux is calculated from the NB and BB fluxes in the following way:
\begin{equation}
\label{eq:flux}
F_{line} = \Delta\lambda_\mathrm{NB} \frac{f_{NB}-f_{BB}}{1-\Delta\lambda_{NB}/\Delta\lambda_{BB}},
\end{equation}
where $\Delta\lambda$ is the width of the NB and BB filters, respectively.

Finally, the $EW$ is calculated as from the NB and BB fluxes:
\begin{align}
\label{eq:EW}
\mathrm{EW} & = \Delta\lambda_\mathrm{NB} \frac{f_{NB}-f_{BB}}{f_{BB}-f_{NB}\left(\Delta\lambda_{NB}/\Delta\lambda_{BB}\right)}
\end{align}
The rest-frame $EW_0$ at the redshift $z$ of the object is:
\begin{equation}
\label{eq:EW0}
\mathrm{EW}_0 = \mathrm{EW}/\left(1+z\right).
\end{equation}

We select as emitters the sources which fulfil the following criteria:
\begin{itemize}
\item $\Sigma>3$: to ensure we select real sources,
\item $EW$ larger than $3$ times the scatter of the BB minus NB colour, in the non-saturated, high S/N regime, to ensure we select real excess sources. The exact cut depends on the cluster, because of the different depths reached in each field.
\end{itemize}
The number of emitters selected is listed in Table~\ref{tab:limmag}.

\subsection{Identifying point sources}\label{sec:stars}

After selecting the emitters, we visually inspect sources to flag potential artefacts as well as any potential star contamination. The number of stars depends heavily on the field, as most clusters are located away from the Galactic plane. However, some clusters (e.g. A545, A2390, `Sausage', `Toothbrush') are located close to the Galactic plane and/or center. Stars with various features in their spectra can contaminate the sample of emitters: in some cases the NB filter can pick up the peak continuum while the BB can have a lot of the absorption, thus mimicking an emission line.

In order to tag an object as a star/point-like object, it has to fulfil any of the following criteria:
\begin{itemize}
\item Classified as star based on spectroscopy: whenever we have a spectroscopically confirmed star we remove it;
\item Classified as star based on morphology: a star is classified as such if we tag it as a star in the visual inspection and it is also unresolved. In order to check  that a source is unresolved we require the source to have a FWHM smaller than the average of the field and well as an ellipticity below $0.2$ in both the NB and the BB filter;
\item Classified as a star because of its IR colours (see Figure~\ref{fig:colcol}): we use the criteria defined in \citet{2012MNRAS.420.1926S}, to select red stars:
\begin{equation}
\label{eq:redstars}
(g-r) > 2 (J-Ks) + 1 \quad \& \quad (g-r)>0.8 \quad  \& \quad (J-Ks) > -0.7
\end{equation}
\item Classified as a blue star or quasar according to the criteria from \citet{2012MNRAS.420.1926S}:
\begin{equation}
\label{eq:bluestars}
(g-r) > 2 (J-Ks) + 1 \quad \& \quad (g-r)<0.8 
\end{equation}
\item Classified as a star because of its optical colours: we use the criteria defined in \citet{2015MNRAS.453..242S}, which removes L and M dwarf stars:
\begin{equation}
\label{eq:optstars}
(g-r) > ( 7/3 (r-i) - 2/3 ) \quad \& \quad (g-r)>1.0
\end{equation}
\end{itemize}

\subsection{Selection of H$\alpha$ candidates}\label{sec:obs:selectionHA}

The sample of potential line emitters is expected to be dominated by H$\alpha$ emitters at the redshifts of the clusters. However, we will also detect other line emitters with shorter intrinsic wavelength, but redshifted at higher z compared to the cluster distance. The most numerous interlopers expected are: H$\beta$ ($\lambda_\mathrm{rest}=4861$\;{\AA}) and [O{\sc iii}]$\lambda\lambda4959,5007$ emitters at $z\sim0.52-0.74$ and [O{\sc ii}] ($\lambda_\mathrm{rest}=3727$\;{\AA}) emitters at $z\sim1.0-1.3$, and to a lesser degree 4000\;{\AA} break galaxies \citep[e.g.][]{2008ApJS..175..128S, 2015MNRAS.453..242S}. 

\begin{figure}
\centering
\includegraphics[trim=0cm 0cm 0cm 0cm, width=0.45\textwidth]{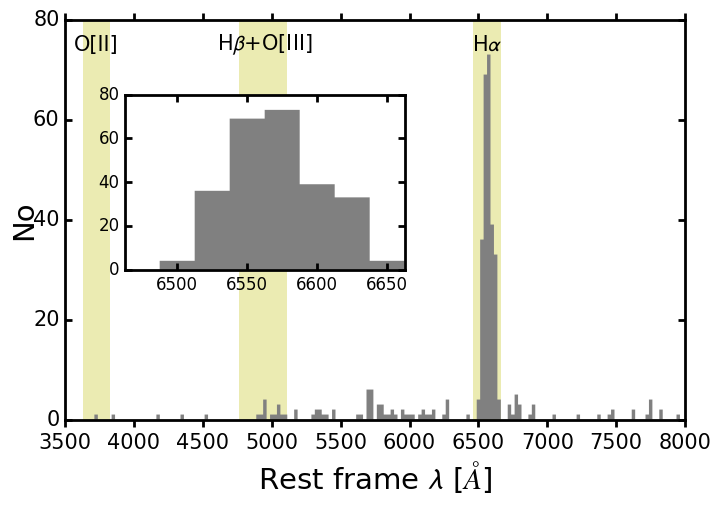}
\vspace{-10pt}
\caption{Histogram of the distribution of emitters with spectroscopic redshift. Because of the different NB filters tracing different redshift ranges, we transform into the rest-frame of the main emission line. The ranges for which H$\alpha$, O[{\sc II}], H$\beta$ and O[{\sc III}] emitters are expected to be picked up by our filters are marked with the shaded areas. The distribution is clearly dominated by H$\alpha$ emitters indicating the filters properly select emitters. Note however, that most of the spectra were targeting the red sequence of the clusters, hence the number of emitters with spectra is rather low. However, the chance of an emitter at the cluster redshift to have a spectrum is still much larger than if they were at a higher redshift, hence there is a bias in source redshifts. The number of spectra for sources at different redshift from the cluster distance is therefore lower than in reality.}
\label{fig:histz}
\end{figure}

Figure~\ref{fig:histz} lists the restframe wavelength of emitters for which we have a spectroscopic redshift. We also overplot the wavelength ranges where given a filter width of $200$\,{\AA} (maximum width of the NB filters we use) would pick up these lines. The H$\alpha$ selection is very good, as exemplified by the clear peak in around the H$\alpha$ wavelength. 

We classify emitters as high-confidence H$\alpha$, uncertain and definitely not H$\alpha$. We can outright remove an emitter if we have spectroscopy confirming it is an emitters at higher redshift (Figure~\ref{fig:histz}). We mark a source as high-confidence H$\alpha$ if it fulfils at least one of these two criteria: i) it has a size of more than $4$\,arcsec on the sky, ii) its spectroscopic redshift is within the redshift range covered by the NB filter. The first criterion was used in \citet{2014MNRAS.438.1377S}, as high-z emitters have a very low chance to reach sizes imposed by a $4$\,arcsec aperture ($10-15$\,kpc size for the redshift range covered by our sources). If these sources were higher redshift, they would be at least $25$\,kpc if they were [O{\sc iii}] emitters at $z\sim0.5$ and $34$\,kpc if they were [O {\sc ii}] emitters at $z\sim1.3$. For many sources we have spectroscopy confirming their H$\alpha$ nature, however this of course does not cover all the sources picked up by the NB filter. However, note the very prominent peak around the H$\alpha$ wavelength for emitters with redshift, which indicates our selection is reliable for H$\alpha$ sources (Figure~\ref{fig:histz}). 

The rest of the sources can either be H$\alpha$ or other line emitteres. With the bands that we possess and the non-uniform data availability and quality for each cluster, it is hard to securely separate H$\alpha$ emitters from other high-z emitters. On a case by case basis, for smaller sources without spectroscopy, we cannot be sure they are H$\alpha$ emitters or other high-z emitters. We therefore follow the statistical method of \citet{2014MNRAS.438.1377S} of using high quality data in deep extragalactic fields to study the fraction of H$\alpha$ emitters in a population of line emitters. We improve on the work from \citet{2014MNRAS.438.1377S} by adding new data from \citet{2015MNRAS.453..242S}. We therefore combine 3 datasets: very deep COSMOS H$\alpha$ NB data at $z\sim0.4$ \citep{2013MNRAS.428.1128S} and at $z\sim0.2$ \citep{2008ApJS..175..128S}, with relatively poor coverage for bright sources, and wide area H$\alpha$ data at $z\sim0.2$ to especially have a better handle of the fractions for bright sources \citep{2015MNRAS.453..242S}. As expected, towards high fluxes (i.e. at bright luminosities) the H$\alpha$ fraction increases fast, as shown in Figure~\ref{fig:fractions}. The functional form for the H$\alpha$ fraction dependence on the luminosity is shown below:
\begin{align}
\label{eq:frac}
\mathrm{frac}_{H\alpha} & = 13.448 \log^4 \left(\frac{L_\mathrm{H\alpha}}{\rm erg\,s^{-1}}\right) -2206.61 \log^3\left(\frac{L_\mathrm{H\alpha}}{\rm erg\,s^{-1}}\right) \notag \\
						& + 1.356\times 10^5 \log^2 \left(\frac{L_\mathrm{H\alpha}}{\rm erg\,s^{-1}}\right) - 3.708 \times 10^6 \log\left(\frac{L_\mathrm{H\alpha}}{\rm erg\,s^{-1}}\right) \notag \\
						& + 3.798 \times 10^7,
\end{align}

When building an H$\alpha$ luminosity function (see Section~\ref{sec:LF}), we apply the fractions derived above to statistically select the appropriate number of H$\alpha$ sources, from the pool of emitters. The number of likely H$\alpha$ emitters, including confident H$\alpha$ sources as well as number of H$\alpha$ obtained by applying the fractions for the rest of the emitters, are listed in Table~\ref{tab:LFs}. The number of emitters in each field can be found in Table~\ref{tab:limmag}. The total number of emitters is $5905$.

\section{H$\alpha$ luminosity function}\label{sec:LHA}

The luminosity function (LF) of H$\alpha$ emitters is obtained by binning emitters depending on their luminosity, diving by the survey volume and fitting with a Schechter function \citep[see Section~\ref{sec:LF},][]{1976ApJ...203..297S} to described the density of emitters. With the goal of building LFs by combining different fields based on cluster properties, we first need to obtain H$\alpha$ fluxes and correct for incompleteness arising from our $EW$ and $\Sigma$ cuts, as well as correct the probed cosmic volumes for the filter profile. These steps are described below.

\subsection{[N{\sc ii}] contamination}
Given the small difference in wavelength, our NB filters will measure the sum of H$\alpha$ and [N{\sc ii}]$_{6450,6585}$. Therefore, the line flux we measure needs to be corrected to obtain H$\alpha$ fluxes. We remove the [N{\sc ii}] contamination from the flux using the relation derived by \citet{2012MNRAS.420.1926S}, in which the [N{\sc ii}] contamination to the flux is a function of $EW$:
\begin{equation}
\label{eq:NII}
f=-0.924+4.802E-8.892E^2+6.701E^3-2.27E^4+0.279E^5,
\end{equation}
where $f$ is the log of the ratio of [N{\sc ii}] to the total flux and $E=\log_{10}(\mathrm{EW}_0(\mathrm{H}\alpha+[\mathrm{N}\textsc{ii}]))$. The mean [N{\sc ii}] contamination is about $30$ per cent of the total blended flux and is consistent with spectroscopy from e.g. \citet{2015MNRAS.450..630S}. This corresponds to roughly sub-solar to solar metallicity sources.

\begin{figure}
\begin{center}
\includegraphics[trim=0cm 0cm 0cm 0cm, width=0.475\textwidth]{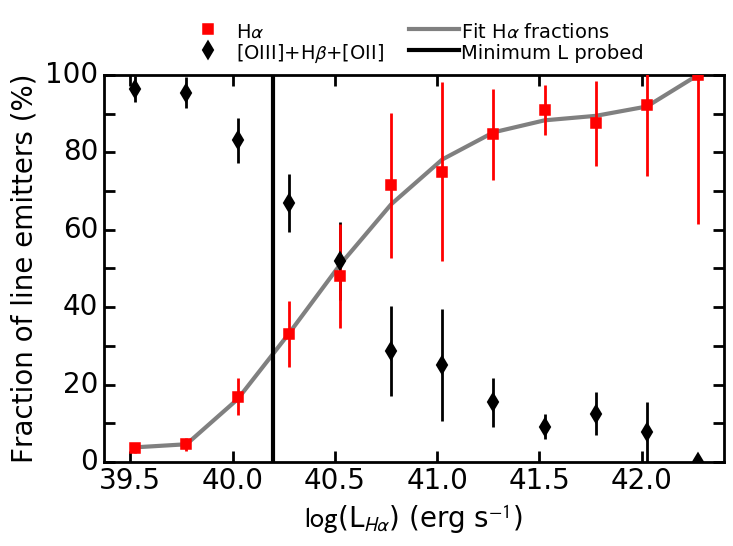}
\end{center}
\vspace{-10pt}
\caption{Fraction of H$\alpha$ emitters expected from a population of line emitters selected with a NB survey, as function of H$\alpha$ luminosity. The gray line displays the H$\alpha$ fraction fit as a function of the luminosity.}
\vspace{-10pt}
\label{fig:fractions}
\end{figure}

\subsection{H$\alpha$ luminosity}
After correcting for the [N{\sc ii}] contamination, we calculate corrected H$\alpha$ fluxes $F_\mathrm{H\alpha}$. The H$\alpha$ luminosity is then defined as:
\begin{equation}
\label{eq:L}
L_\mathrm{H\alpha}=4 \pi D^2_{L}(z) F_\mathrm{H\alpha},
\end{equation}
where $D_{L}(z)$ is the luminosity distance of each cluster (see Table~\ref{tab:obs}).

\subsection{Completeness correction}
At faint luminosities or low $EW$, our survey we will only recover a fraction of the true number of sources. We correct for incompleteness by selecting random subsamples of sources consistent with being non-emitters and adding increasing larger line fluxes to their fluxes. We then pass the fake emitters through the same selection criteria as the real sources (see Section \ref{sec:selection}). We perform the study independently for each sources and each individual CCD to test how many sources we recovered as function of luminosity. At each luminosity, we correct the LF for incompleteness. We refer the readers to \citet{2012MNRAS.420.1926S} and \citet{2013MNRAS.428.1128S} for further details on the method.

\subsection{Filter correction}

Table \ref{tab:obs} lists the expected cosmic volumes probed in each field, taking into account the effective area covered by the camera on sky, after masking bright stars and noisy regions. The volumes vary with the FWHM of the NB filters as well as the redshift of H$\alpha$ we are tracing in each field. The volumes are initially calculated assuming the NB filters have a perfect top-hat (TH) shape with a FWHM as stated in Table \ref{tab:filters}. However, the actual shape of the filters deviates from a TH (see Figure \ref{fig:filters}), which means not all sources located in the wings of the filter will be detected. Following the method described in \citet{2009MNRAS.398...75S} and \citet{2012MNRAS.420.1926S}, we correct the LF for the shape of the filters to take into account the sources missed at the edged of the filter. For each field and filter, we generate a sample of H$\alpha$ emitters as would be selected by a perfect TH filter and bin them according to luminosity. We compute a first pass LF fit by a Schechter function. We then generate an idealised sample of H$\alpha$ emitters according to the Schechter function just derived. We then pass this idealised population through the real filter profile to study the recovery rate of emitters at each wavelength covered by the filter.

\subsection{Survey limits}\label{sec:LHA:limits}
At $50$ per cent completeness, the average limiting H$\alpha$ luminosity varies between $10^{40.2-41.3}$\,erg\,s$^{-1}$ (for full details see Table~\ref{tab:limmag}). This is driven by the depth of the observations as well as the redshift of the sources. Assuming the \citet{1998ARA&A..36..189K} relation, corrected for a Chabrier IMF, this corresponds to limiting SFRs of $0.07-0.78$ $M_{\odot}$ yr$^{-1}$, when no intrinsic dust extinction is applied. This corresponds to $0.03-0.3$\,$SFR^*$ at the respective redshifts of the clusters, with the average being $0.1$ $SFR^*$. 

\subsection{H$\alpha$ luminosity function}\label{sec:LF}

We bin the emitters based on luminosity, corrected for [N{\sc ii}] contamination, and add their associated inverse volume to obtain LFs. We only add sources in volumes with at least $50$ per cent completeness. As mentioned in Section \ref{sec:obs:selectionHA}, we count the sources we are confident are H$\alpha$ emitters with a weight of 1 and we apply a statistical H$\alpha$ probability fraction for sources we cannot be sure are H$\alpha$ and not higher-z sources. We correct the LFs for incompleteness and for the filter profile, but note that we are not correcting for intrinsic dust extinction. 

We use a least-squares fit to parametrise the binned data with a \citet{1976ApJ...203..297S} function, using Poissonian errors:
\begin{equation}
\label{eq:schechter}
\phi(L_{\mathrm{H}\alpha}) \mathrm{d} L_{\mathrm{H}\alpha} = \phi^*\left(\frac{L_{\mathrm{H}\alpha}}{L^*_{\mathrm{H}\alpha}}\right)^{\alpha} e^{-\frac{L_{\mathrm{H}\alpha}}{L^*_{\mathrm{H}\alpha}}} \mathrm{d} \left(\frac{L_{\mathrm{H}\alpha}}{L^*_{\mathrm{H}\alpha}}\right),
\end{equation}
where $\phi^*$ is the typical number density of H$\alpha$ sources, $L^*_{\mathrm{H}\alpha}$ is the characteristic luminosity and $\alpha$ is the faint-end slope of the LF. We allow for all three parameters of the fit to vary freely. We perform the fit using a range of different $\log L$ bins: with widths $\Delta \log L$ from $0.15$ to $0.4$ and starting bins $\log L_\mathrm{min}$ ranging from $40.$ up to $40.5$. 

We tested a number of different ways to bin the data in order to avoid reporting parameters which could be biased by a particular binning choice. We first binned the data with a random choice of bin widths and bin centres and fit a LF to all the resampled data. Secondly, we also rebinned these resamples to a wider L grid and fit an average LF. We also fit individual LFs to each of our random choices of bin width and calculated the average of the results. 

We also tested fits with all three parameters free and found that in many cases the overall fit was biased because of the faint-end slope. To test the robustness of the fits with $\alpha$, $\phi^*$ and $L^*$ free, we studied the faint-end by fitting a straight line to only the faintest bins and found that in some cases this did not match the $\alpha$ obtained by fitting a full LF to all the data. 

\renewcommand{\arraystretch}{1.3}
\begin{table*}
\begin{center}
\caption{Parameters of LFs for different H$\alpha$ samples, with faint-end slope $\alpha=-1.35$. These were selected inside and outside the clusters, and in clusters of different relaxation states, masses, luminosities and redshifts. We also list the total volume of each of the combined LFs. The last column lists the number of likely H$\alpha$ emitters, including secure H$\alpha$ emitters confirmed through spectroscopy or their size. For the rest of the emitters we applied the fractions derived in equation~\ref{eq:frac}.}
\vspace{-5pt}
\begin{tabular}{l c c c c}
\hline\hline
Stack & $\log \phi^*$ (Mpc$^{-3}$) & $\log L^*_\mathrm{H\alpha}$ (erg s$^{-1}$)  & V [$10^4$ Mpc$^3$] & H$\alpha$ sources \\ \hline
\citet{2015MNRAS.453..242S} fit & $-2.85 \pm 0.03$ & $41.71 \pm 0.02$ \\ \hline
All cluster fields & $-1.95^{+0.06}_{-0.06}$ & $41.56^{+0.05}_{-0.05}$ & \hspace{-4pt}13.0 & 3472 \\ \hline
inside $2$\,Mpc & $-1.98^{+0.06}_{-0.04}$ & $41.56^{+0.03}_{-0.03}$ & 4.4 &  1203\\
beyond $2$\,Mpc & $-1.92^{+0.04}_{-0.06}$ & $41.53^{+0.05}_{-0.05}$ & 8.6 & 2211 \\ \hline
Merging, $0-0.5$\,Mpc    & $-1.59^{+0.06}_{-0.07}$ & $41.68^{+0.09}_{-0.06}$ & 0.2 & \phantom{00}92  \\
Relaxed, $0-0.5$\,Mpc    & $-1.92^{+0.07}_{-0.11}$ & $41.71^{+0.09}_{-0.06}$ & 0.1 & \phantom{00}49  \\

Merging, $0.5-1.0$\,Mpc  & $-1.71^{+0.06}_{-0.04}$ & $41.56^{+0.03}_{-0.03}$ & 0.4 & \phantom{0}174  \\
Relaxed, $0.5-1.0$\,Mpc  & $-2.04^{+0.03}_{-0.05}$ & $41.50^{+0.01}_{-0.01}$ & 0.5 & \phantom{00}94  \\

Merging, $1.0-1.5$\,Mpc  & $-1.77^{+0.06}_{-0.06}$ & $41.56^{+0.05}_{-0.03}$ & 6.0 & \phantom{0}248 \\
Relaxed, $1.0-1.5$\,Mpc  & $-2.31^{+0.06}_{-0.09}$ & $41.62^{+0.09}_{-0.05}$ & 6.8 & \phantom{0}107 \\

Merging, $1.5-2.0$\,Mpc  & $-1.92^{+0.09}_{-0.09}$ & $41.50^{+0.08}_{-0.06}$ & 8.0 & \phantom{0}241 \\
Relaxed, $2.0-2.0$\,Mpc  & $-2.28^{+0.04}_{-0.06}$ & $41.71^{+0.08}_{-0.06}$ & 8.9 & \phantom{0}159 \\

Merging, outside $2.0$\,Mpc  & $-1.86^{+0.05}_{-0.04}$ & $41.53^{+0.05}_{-0.03}$ & 46.6 & 1574 \\ 
Relaxed, outside $2.0$\,Mpc  & $-2.13^{+0.04}_{-0.05}$ & $41.53^{+0.06}_{-0.06}$ & 33.3 & \phantom{0}599 \\ \hline

Mergers & $-1.80^{+0.06}_{-0.07}$ & $41.53^{+0.05}_{-0.03}$ & 1.9 & \phantom{0}755 \\
Relaxed & $-2.16^{+0.06}_{-0.06}$ & $41.59^{+0.05}_{-0.05}$ & 2.1 & \phantom{0}409 \\
Relics &  $-1.53^{+0.06}_{-0.09}$ & $41.41^{+0.03}_{-0.01}$ & 1.0 & \phantom{0}430 \\
Haloes or haloes and relics & $-1.80^{+0.05}_{-0.06}$ & $41.53^{+0.03}_{-0.01}$ & 1.6 & \phantom{0}632\\ \hline
Low z ($0.15<z<0.20$) & $-1.98^{+0.06}_{-0.06}$ & $41.59^{+0.06}_{-0.03}$ & 0.8 & \phantom{0}292 \\
Mid z ($0.20<z<0.25$) & $-2.28^{+0.04}_{-0.08}$ & $41.65^{+0.07}_{-0.05}$  & 2.9 &  \phantom{0}574 \\
High z ($0.25<z<0.31$) & $-1.17^{+0.04}_{-0.06}$ & $41.32^{+0.03}_{-0.01}$ & 0.7 & \phantom{0}337 \\ \hline
Low mass$^\ddagger$ ($4\times10^{14} M_{\sun}<M<10\times10^{14} M_{\sun}$) & $-2.58^{+0.09}_{-0.10}$ & ${41.86}^{+0.15}_{-0.08}$ & 2.1 & \phantom{0}388 \\
High mass ($10\times10^{14} M_{\sun}<M<29\times10^{14} M_{\sun}$) & $-1.89^{+0.04}_{-0.03}$ &  ${41.53}^{+0.03}_{-0.03}$ &  2.0 & \phantom{0}776 \\ \hline
Low L ($5\times10^{44}$\,erg\,s$^{-1}<L_\mathrm{X}<10\times10^{44}$\,erg\,s$^{-1}$) & $-2.22^{+0.11}_{-0.12}$ & ${41.68}^{+0.19}_{-0.10}$ &  2.1 & \phantom{0}425 \\ 
High L ($10\times10^{44}$\,erg\,s$^{-1}<L_\mathrm{X}<38\times10^{44}$\,erg\,s$^{-1}$) &  $-1.77^{+0.04}_{-0.05}$ & ${41.50}^{+0.03}_{-0.03}$ & 1.8 & \phantom{0}755 \\
\hline
\end{tabular}
\label{tab:LFs}
\end{center}
{\small $\dagger$ Faint end slope was fixed to value derived in \citet{2008ApJS..175..128S}.\\
$\ddagger$ A1300 was removed from the stack as it was dominating the LF fit.}
\end{table*}
\renewcommand{\arraystretch}{1.1}

In order to further test this, we also performed a resampling analysis, where for each combined volume, we removed one-by-one each cluster from the stack, to see whether a particular cluster is dominating the fit. We discovered that the fits were not robust when removing a cluster from the fits, and the LF fits to these data, while consistent within the error bars, were in many cases at the very edge of inconsistency to the LF obtained using all the clusters in the `stack'. Additionally the error bars on each LF parameter were large. We conclude that we can not derive a very robust faint-end slope value. This is mostly driven by the depth of our data. Additionally, when combining different clusters, at the very faintest bins, the combined LF is dominated by a few clusters, which might bias the results. We therefore decided to fit LFs by fixing $\alpha$ to values derived from deep data, specifically $-1.35$ from \citet{2008ApJS..175..128S} and $-1.7$ from \citet{2007ApJ...657..738L}. We find that our LF parameters have lower errors and are more robust against removing individual clusters from the combined volume when using the flatter faint-end slope $-1.35$.

We also noticed that at the very brightest luminosities, beyond $10^{42.2}$ erg s$^{-1}$, there was a very high bin, inconsistent with the usual drop of the LF towards these luminosities. This is caused by $<5$ sources above the expected Poissonian variation. Even though these have passed visual inspection, they are compact sources and hence they could be AGN. We will follow up this sources and inspect their nature in a future paper. For the present study, in order to make out fits more robust, we are not considering bins with $L>42.2$ in our LF fits.

Overall, after fixing the faint end and removing the very bright luminosity bins, we find that all the methods we used to bin the data and fit LFs produce results which are consistent within the error bars. In general, the individual binning choices also agree with the average fits within the errors, with the exception of a limited number of binning choices, as expected. We finally bin all the $\phi$ values obtained with a range of bin widths and bin centres to produce an average binning. We calculate the error as the standard deviation of the $phi$ values falling within each final bin. We therefore report the LF parameters resulting from a binning which reproduced well the average LF and also results in LF parameters with small errors, again indicating a good fit.

\section{Results}\label{sec:results}

Our main goal for this work is to contribute to our understanding of the drivers of SF in clusters. In order to do so, we need to compare relaxed and merging clusters, look for any trends with mass and/or luminosity and of course compare to results obtained over wide areas to quantify the statistical behaviour of the Universe in lower density environments. Therefore, we bin the emitters based on a number of criteria, according to the cluster properties listed in Table~\ref{tab:clusters}. \vspace{10pt}

\begin{itemize}
\item {\bf General, all fields sample}: we bin all the emitters for all the fields, both inside and outside of the clusters;\vspace{5pt}
\item {\bf Environment}:
	\begin{itemize}\vspace{-5pt}
	\item Clusters: we stack all the emitters within clusters out to a projected radius of 2\,Mpc away from the cluster centre;
	\item Field around clusters: stack of emitters located around the cluster, more than 2\,Mpc away from the cluster centre;\end{itemize}
\item {\bf Merger state:}
	\begin{itemize}\vspace{-5pt}
	\item Relaxed clusters: stack of all the relaxed clusters;
	\item Merging clusters: clusters that host relics and clusters that host haloes; 
	\end{itemize}
\item {\bf Mass of the cluster} -- bin emitters within clusters of certain mass (low and high mass);
\item {\bf Luminosity} -- bin cluster emitters depending on the host cluster X-ray luminosity (low and high X-ray luminosity);
\end{itemize}

\begin{figure}
\begin{center}
\includegraphics[trim=0cm 0cm 0cm 0cm, width=0.475\textwidth]{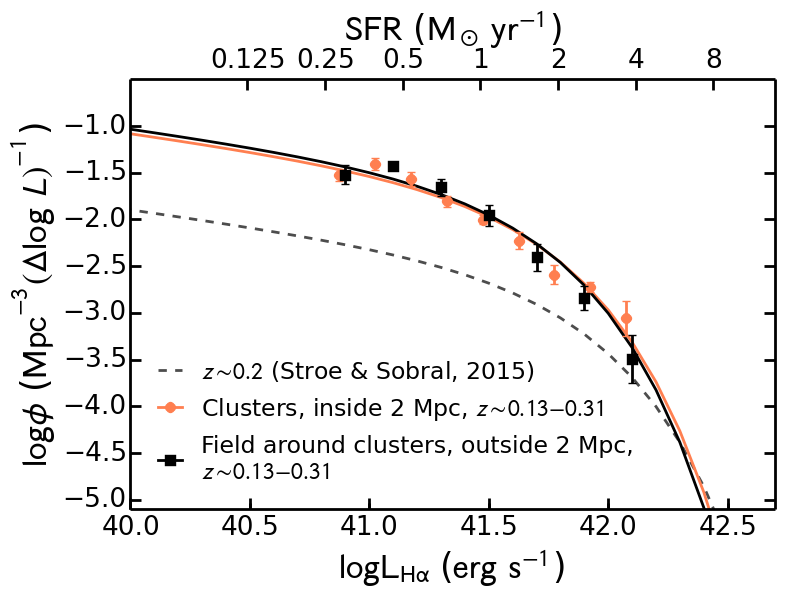}
\end{center}
\vspace{-10pt}
\caption{The H$\alpha$ LF averaged over all the clusters in our sample, within 2\,Mpc of the cluster centre. The cluster LF is consistent with the larger scale environment of the clusters beyond 2\,Mpc. However, compared to an average field \citep{2015MNRAS.453..242S}, the clusters reside in overdense environments. Note that the H$\alpha$ properties inside and outside of the clusters is very difference for merging and relaxed clusters (see Figure~\ref{fig:environ}).}
\vspace{-10pt}
\label{fig:LFenviron}
\end{figure}

This information is also summarised in Table~\ref{tab:LFs}, where we also list the best fit Schechter function parameters, volumes of each combined dataset and the number of H$\alpha$ emitters used to produce the LFs. As reference for these LFs, we use the $z\sim0.2$ H$\alpha$ LF derived in \citet{2015MNRAS.453..242S}, which combines deep data within a smaller FOV \citep{2008ApJS..175..128S} to capture faint H$\alpha$ emitters, with shallower data over a large field to overcome cosmic variance and capture rare, bright emitters. Our survey sits at a third of the volume from \citet{2015MNRAS.453..242S} ($\sim1.3\times10^5$\,Mpc$^3$ compared to the larger $3.5\times10^5$\,Mpc$^3$), albeit our H$\alpha$ emitters are selected over a wider redshift spread. A typical deep field NB H$\alpha$ survey such as the one of \citet{2008ApJS..175..128S} covers $\sim3.1\times10^4$ Mpc$^3$. Most of our combined datasets reach volumes of $50-75$ per cent of that value. The smallest volumes are for the combined cluster cores, as well as for the low-z data, as expected. The small volumes will of course mean our combined volumes are highly sensitive to cosmic variance, as is already exemplified by the differences in the numbers of H$\alpha$ emitters in each combined dataset. For example, for similar volumes, the mergers combined dataset has a factor of $\sim1.8$ more H$\alpha$ sources than the relaxed cluster data. However this enables us to investigate environmental trends.

\subsection{Environment}

Compared to the large H$\alpha$ survey from \citet{2015MNRAS.453..242S} which covers voids, rich and dense fields and greatly overcomes cosmic variance, we find that, statistically, the cluster fields targeted in this survey live in generally rare/extreme, overdense environments (at $>27\sigma$ level). Otherwise, the characteristic luminosity is in agreement (difference of less than $3\sigma$). This is interesting since \citet{2015MNRAS.453..242S} used both very deep data over a small field as well as a very large, shallow survey to obtain the LF. \citet{2015MNRAS.453..242S} predict that from cosmic variance, our $\phi^*$ and $L^*$ of sources should be within $25-30$ per cent from numbers obtained in a very large survey. Our survey is right at the edge of this prediction, which is expected given the survey is targeting the densest parts of the Universe.  

On average, the LF in the cluster sample (within a projected radius of 2\,Mpc of the cluster centre) is similar to that in the field around them (outside 2\,Mpc from the cluster centre). This is illustrated in Figure~\ref{fig:LFenviron}. However, this average shape is obtain by averaging between the opposing behaviours of the relaxed and merging clusters.

\begin{figure}
\begin{center}
\includegraphics[trim=0cm 0cm 0cm 0cm, width=0.475\textwidth]{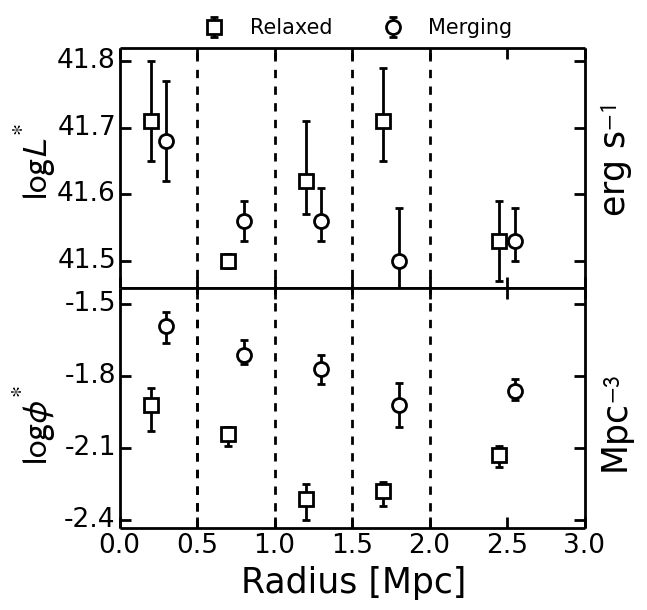}
\end{center}
\vspace{-10pt}
\caption{Dependence of the H$\alpha$ LF parameters on environment. Points have been shifted in projected radius for clarity. There are striking differences between the behaviour of relaxed and merging clusters in both $\phi^*$ and $L^*$. The emitters are binned in five regions, indicated with the vertical dashed lines: a circle within 0.5\,Mpc from the centres of each cluster, an annulus between 0.5 and 1\,Mpc radius, another annulus between 1\,Mpc and 2\,Mpc and then all emitters outside of 2\,Mpc from the cluster centre.}
\vspace{-10pt}
\label{fig:environ}
\end{figure}

\begin{figure*}
\begin{center}
\includegraphics[trim=0cm 0cm 0cm 0cm, width=0.475\textwidth]{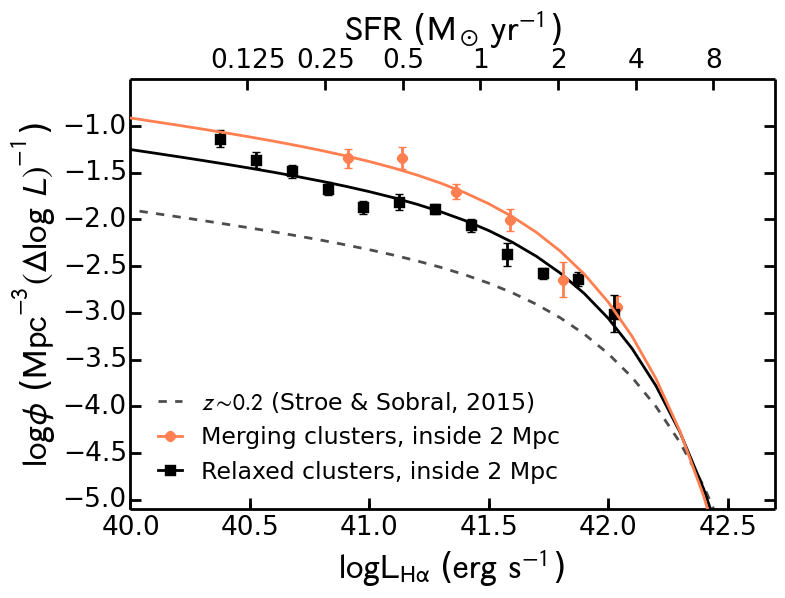}
\includegraphics[trim=0cm 0cm 0cm 0cm, width=0.475\textwidth]{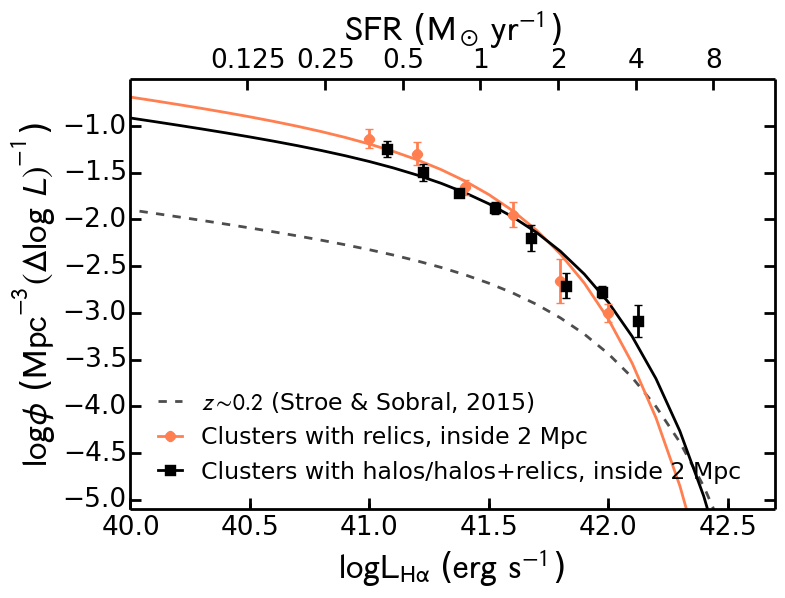}
\end{center}
\vspace{-10pt}
\caption{Left panel: The H$\alpha$ LF for merging and relaxed clusters (within 2\,Mpc of centre). Note the differences in normalisation of the two LFs. Right panel: The H$\alpha$ LF for clusters hosting relics and haloes, respectively (within 2\,Mpc of centre). The presence of shocks mildly boosts the number of H$\alpha$ emitters compared to a cluster hosting a halo.}
\vspace{-10pt}
\label{fig:LFdiffuse}
\end{figure*}

However, when dividing in smaller annular regions, we find trends with cluster-centric projected radius (see Figure~\ref{fig:environ}). The dependence with radius of the LF parameters differs between relaxed and merging clusters. While both have similar $L^*$ values at the core and outside the clusters, the trends between 0.5 Mpc and 2.0 Mpc are vastly different. While for relaxed clusters the characteristic luminosity slowly rises across this region, for merging clusters it systematically drops from the cluster core to the to the value in the field outside the cluster (beyond 2 Mpc from the cluster centre). In terms of characteristic density, there is a trend of dropping $\phi^*$ from cores to outskirts ($1.5-2$\,Mpc). However, for the relaxed clusters $\phi$ drops below the field value for the cluster outskirts, for the merging cluster no regions falls below the field around the clusters. Overall, every region within merging clusters is denser in H$\alpha$ emitters when compared to the densest region (the core) of relaxed clusters.

\subsection{Relaxation state}

Merging environments are different from relaxed clusters: merging clusters have a higher $\phi^*$ (at $4\sigma$ level) and are overdense in H$\alpha$ emitters at all luminosities, compared to relaxed cluster fields (see left panel of Figure~\ref{fig:LFdiffuse}). As mentioned before, the merging clusters are on average more massive than the relaxed cluster, however the different average cluster mass of the merging and relaxed samples cannot explain the differences we see, as will also be shown in Section~\ref{sec:mass}.

\subsection{Presence of shocks and turbulence}

Clusters hosting shocks, have on average, higher characteristic H$\alpha$ densities and lower $L^*$ compared to those hosting turbulence, marked by the presence of radio haloes (see right panel of Figure~\ref{fig:LFdiffuse}). The differences are significant at the $\sim3-4\sigma$ level. 

There seems to be evidence for a decreasing $L^*$ and increasing $\phi^*$ from relaxed, to halo-hosting clusters to relic hosting clusters. This indicates that relic clusters host numerous H$\alpha$ emitters fainter than the characteristic luminosity. By contrast, high luminosity emission might be suppressed. 

\subsection{Cluster mass and X-ray luminosity dependence}\label{sec:mass}

An important aspect is that the high L combined volume contains both relaxed and merging objects in equal numbers (4 relaxed and 4 merging for the high-L `stack'). However, the high mass combined volumes are dominated by disturbed objects (8 merging vs 3 relaxed), so, with our sample, we cannot fully disentangle the effects of mass and relaxation state.

\begin{figure*}
\begin{center}
\includegraphics[trim=0cm 0cm 0cm 0cm, width=0.475\textwidth]{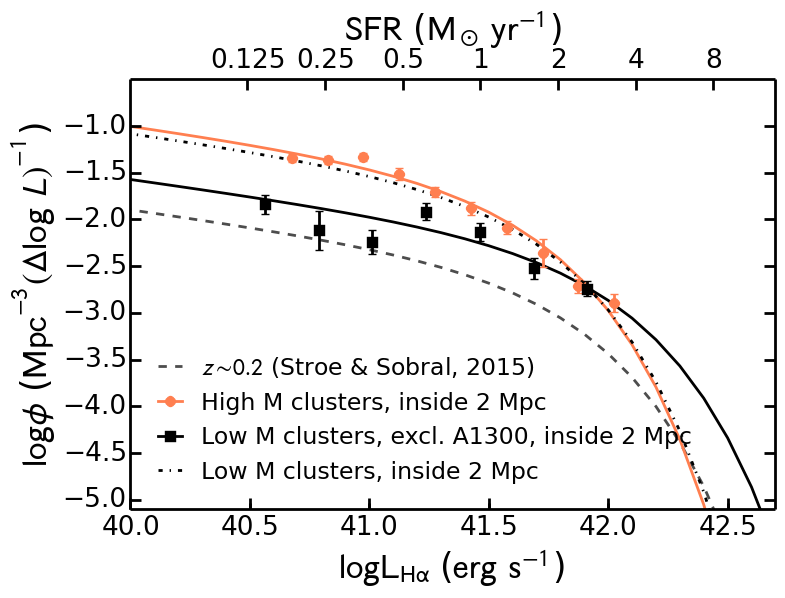}
\includegraphics[trim=0cm 0cm 0cm 0cm, width=0.475\textwidth]{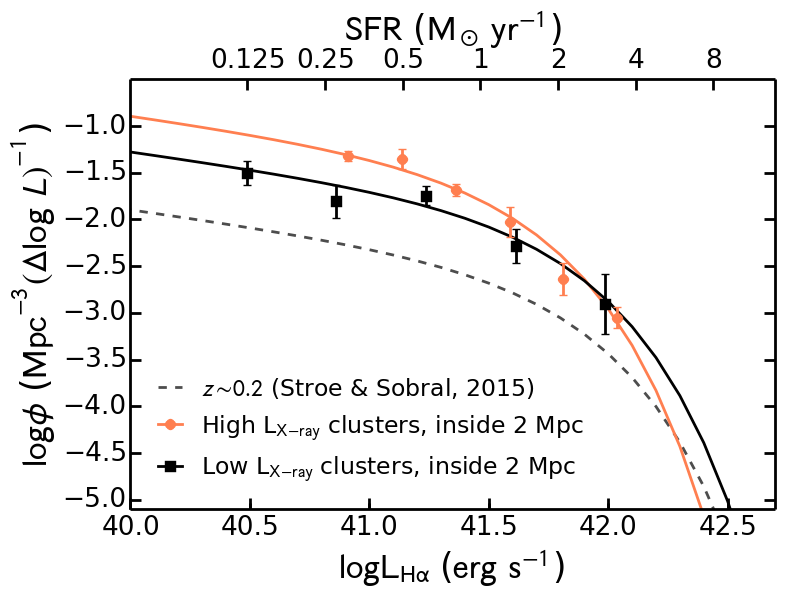}
\end{center}
\vspace{-10pt}
\caption{The H$\alpha$ LF for stacks of low and high mass cluster (left panel) and low and high luminosity clusters, respectively (right panel). Only the emitters within 2\,Mpc of the cluster centre were considered. There is marginal evidence for X-ray bright clusters hosting more numerous, lower luminosity H$\alpha$ emitters on average. There are no differences between the low and high mass cluster LF, when A1300 is included in the low-M `stack'.}
\vspace{-10pt}
\label{fig:LFL}
\end{figure*}

High X-ray luminosity clusters are overdense in H$\alpha$ emitters compared to low-L clusters (see right panel of Figure~\ref{fig:LFL}). The differences between $\phi^*$ are $>3\sigma$ away. The $L^*$ for the two samples are consistent within the error bars.

In the case of low and high mass clusters, we observed no significant differences in the shape of the LF. However, once we remove cluster A1300, which has a low mass, there are statistically significant difference between the low and high mass stacks in both the $\phi^*$ and the $L^*$ (left panel, Figure~\ref{fig:LFL}). The low-mass combined volume is the only case where we find a single cluster to dominate the fit. We also note that the average mass of the high-mass combined volume is higher than the average mass of the merging cluster stack ($18$ vs $16$ $10^{14} M_{\sun}$). However, the $\phi^*$ of the merging cluster stack is higher than that of the high mass stack, indicating mass alone cannot explain the increased characteristic density of H$\alpha$ emitters.

\section{Discussion}\label{sec:discussion}

Relaxed clusters have a high density of galaxies compared to the fields around them. In stark contrast to field environments, passive galaxies represent a large fraction of the cluster population. Galaxies in $z=0$ relaxed clusters are thought to have formed most of their stars in a single burst of SF at large cosmic time and then evolved passively without the possibility of accretion of new material. The hot ICM of relaxed clusters also has a profound influence on the fate of infalling galaxies: ram pressure stripping and other interactions may lead to the removal of gas, thus accelerating the evolution of field spirals into passive cluster ellipticals or S0s. 

However, disturbed clusters have not been explored as much and they offer tantalising opportunities to study environments and effects quite similar to high-redshift proto-clusters. To test whether the SF properties of merging cluster galaxies are different from those in relaxed clusters, we are exploring our sample of $\sim20$ clusters which span a range in mass, luminosity and redshift. Our goal is to find the main driver of SF and transformation of gas-rich spirals into gas-poor ellipticals in disturbed environments and their larger-scale surroundings.

We find that both relaxed and merging clusters and their larger scale structure are overdense in H$\alpha$ emitters when compared to an average cosmic volume. One might expect that the very low fraction of spiral galaxies in clusters would lead to a lower H$\alpha$ LF normalisation compared to an average field volume. Nevertheless, this seems not to be the case most probably because of the large overdensities of galaxies clusters represent. Clusters reside in a large web of filaments, which have been found to be rich in star-forming galaxies \citep{2014ApJ...796...51D}. Our results support this scenario.  

We also study the differences between relaxed and merging clusters, also separating into disturbed clusters hosting shocks (using relics as proxy) and those hosting increased turbulence (using haloes as proxy). The H$\alpha$ properties of galaxies within relaxed and merging clusters are different. The $\phi^*$ and $L^*$ vary in different ways with cluster centric distance for the two classes of clusters. 

In the cores of both relaxed and merging clusters there seems to be a peak in characteristic luminosity and density. Important transformations can happen in the densest parts of the ICM: at cluster cores we could be seeing an increase in AGN activity and in galaxy-galaxy mergers.

For relaxed clusters, the characteristic densities of H$\alpha$ emitters drop from core to immediate outskirts, where they fall below the field value around the clusters. This might indicate a suppression of SF in a fraction of infalling galaxies. However, the $L^*$ of galaxies located towards relaxed cluster outskirts is higher than the field, indicating that in the galaxies surviving the infall into the cluster, there might be triggered SF. This could be caused for example by ram-pressure stripping. 

Merging clusters have a high $\phi^*$, throughout the cluster volume, at all radii, staying always above the field levels. Therefore, merging clusters clearly present a very different environment from relaxed clusters. In disturbed clusters, the characteristic H$\alpha$ number density is at least as high as the fields. The origin of these galaxies could either be recently accreted field spirals or triggering via ram pressure processes of infalling galaxies, however, we do not find any particular enhancement at at the cluster outskirts. \citet{1983ApJ...270....7D} found star-bursting signatures in spectra of intermediate-redshift cluster galaxies, which they interpreted as ``ram-pressure induced SF": the galaxies were shocked into an increased SFR, before the truncation of SF occurs. This scenario was later confirmed through simulations by \citet{2003ApJ...596L..13B}. 

However, the enhancement in terms of numbers of H$\alpha$ emitters prevails towards the ``cores" of disturbed clusters. Note that part of the emitters may be located towards the outskirts of the clusters, but seen in projection, however this cannot fully explain the increase in H$\alpha$ number density towards the cluster centre. The general picture of massive galaxy clusters involves galaxy cluster populations undergoing a single massive burst of SF at high look back times \citep[e.g.][]{1962ApJ...136..748E,1967ApJ...147..868P, 2003Natur.425..264S, 2010ApJ...709..512R}. Clusters would then grow by mergers with other relaxed clusters hosting predominantly passive galaxies, and by accretion of smaller, more disturbed clusters hosting a larger fraction of spirals as well as field galaxies. The presence of active, H$\alpha$ emitters deep in the core of disturbed clusters in our sample could indicate that the progenitors of the mergers were not relaxed, hence environmental quenching has not been operating for significant amounts of time. However this scenario fails to explain how the merging clusters we are studying have grown to be so massive, if the progenitors were also of young age, hence did not have a lot of time to grow their mass. 

Another scenario would be that the merger induced processes which acted as a catalyst for SF. \citet{2014MNRAS.443L.114R} adapted the ideas from \citet{1983ApJ...270....7D} and \citet{2003ApJ...596L..13B} into simulations where not necessarily ram pressure, but large scale, low Mach number cluster shocks would traverse gas-rich galaxies. They found that indeed such shocks would be capable of triggering SF in cluster galaxies. This could be similar to cold gas streams proposed to fuel the growth of galaxies by penetrating shock-heated media of massive dark matter halos \citep{2009Natur.457..451D}. Additionally these massive clusters could have accreted smaller, young subgroups as well as collapsed filaments which in combination with triggered SF could explain the increased H$\alpha$ density.

Our results also indicate that merging clusters hosting haloes are closer in terms of SF properties to relaxed clusters than relic clusters ($\phi^*$ drops from relic, to halo cluster to relaxed clusters). This could indicate that either halo-hosting clusters are more dynamically evolved than relic clusters, which would allow the galaxies to further evolve into passive galaxies, explaining the deficit of mid-L H$\alpha$ emitters. This is line with theory of diffuse radio emission, which indicates that the halo effect appears later than relics and is a more long lasting one \citep{2014IJMPD..2330007B}. Additionally, turbulence, if indeed correlated with haloes, might not have as much of an effect on SF as shocks. The large scale flows cascade into small scale turbulence on scales of $10-100$\,kpc, thought to cause particle acceleration and hence radio haloes. A possibility is that the turbulence does not penetrate into the intra-galactic medium {and thus is not able} to change galaxies' SF properties.

We find that cluster X-ray luminosity correlates more strongly with SF properties rather than cluster mass. Cluster mass cannot solely explain the evolution with relaxation state in the shape of the H$\alpha$ LF. Naively, we would expect that massive clusters host a larger number of galaxies and assuming a fixed fraction of SF galaxies, hence a larger number of H$\alpha$ emitters. However, cluster mass does not correlate well with cluster X-ray luminosity (Figure~\ref{fig:clusters}), most probably because many of these clusters are not in hydrostatic equilibrium. Therefore, our results indicate the merger status of the host cluster plays an important role in setting the SF trends of cluster galaxies.

\section{Conclusions}\label{sec:conclusion}

We performed an H$\alpha$ narrow band survey of a sample of $19$ clusters with redshifts covering the $0.15-0.31$ range. We selected $>3000$ likely H$\alpha$ emitters over a total volume of $\sim1.3\times10^5$ Mpc$^3$, located in a variety of environments. The H$\alpha$ emitters are located in relaxed and merging clusters of low and high mass and luminosity, as well as in the large scale environment of the clusters. 

With our data, we are studying the effects of environment on the properties of the H$\alpha$ luminosity function, specifically focusing on the way disturbed clusters can drive the SF properties of their members. We also compare relaxed cluster environments to clusters with evidence for large shock waves and increased ICM turbulence. Our main results are:

\begin{itemize}
\item We build a first `universal' H$\alpha$ luminosity function for clusters and their nearby environments. The luminosity function is fit by a Schechter function with fixed $\alpha=-1.35$ and parameters $\log \phi^*=-1.95^{+0.06}_{-0.06}$ and  $\log L^*=41.56^{+0.05}_{-0.05}$. Cluster fields are overdense in H$\alpha$ emitters compared to an average cosmic volume.
\item There is a significant difference between the properties of the H$\alpha$ luminosity function in relaxed and merging clusters, which cannot be solely attributed to the mass of the hosting clusters. The dependence of the LF parameters on cluster centric distance is different for merging and relaxed clusters. 
\item At all projected cluster-centric radii, $\phi^*$ is much higher for merging clusters than for relaxed objects. Merging clusters, especially those with ICM shocks, have a density of H$\alpha$ emitters slightly larger than the field around them.
\item For merging clusters, $L^*$ drops slowly from cores to the field value just outside the cluster, while for relaxed clusters $L^*$ increases towards cluster outskirts. 
\item We speculate that increased AGN activity and galaxy-galaxy mergers can elevate $L^*$ and $\phi^*$ in the cluster cores. At the outskirts of relaxed clusters, accretion of gas rich galaxies can lead to an increase of the typical $L^*$. In merging clusters, triggered SF can occur through interactions with the ICM, cluster-wide shocks. The SF can also be increased through collapsed spiral-rich filaments and accretion of young galaxy groups. 
\item X-ray luminosity, which is related to both mass and merger state of the cluster, seems to have a higher impact of the H$\alpha$ luminosity function than the mass alone. This corroborates the above results that the merger state of the host cluster has a high impact on the SF properties of cluster galaxies.
\end{itemize}

\section*{Acknowledgements}
We thank the referee for the comments which improved the clarity of the paper. We are grateful to W. Dawson and N. Golovich for sharing their published data with us and H. R\"ottgering and D. Wittman for useful discussions. We thank B. Miranda Ocejo, S. Harish, S. Perez, J. Cairns, S. Dempsey and R. Kaiser for their help with the observations. DS acknowledges financial support from Lancaster University through an Early Career Internal Grant A100679, the Netherlands Organisation for Scientific research (NWO), through a Veni fellowship, and from FCT through a FCT Investigator Starting Grant and Start-up Grant (IF/01154/2012/CP0189/CT0010). Based on observations made with the Isaac Newton Telescope (proposals I16AN002, I15AN001, I12BN003, I13BN006) operated on the island of La Palma by the Isaac Newton Group in the Spanish Observatorio del Roque de los Muchachos of the Instituto de Astrof{\'i}sica de Canarias. The research leading to these results is part based on observations takes at the MPG 2.2m telescope, through the OPTICON programme (ID 14B039). OPTICON is supported by the European Community's Seventh Framework Programme (FP7/2013-2016) under grant agreement number 312430. We would like to thank M. Balogh, R. Barrena, W. Boschin, D. Coe, D. Frye, M. Girardi, R. Houghton, F. La Barbera, L. Lemonon, D. Lenze, M. Owers and M. Pierre for making their spectroscopic and photometric catalogues public. Based on observations obtained with MegaPrime/MegaCam, a joint project of CFHT and CEA/IRFU, at the Canada-France-Hawaii Telescope (CFHT) which is operated by the National Research Council (NRC) of Canada, the Institut National des Science de l'Univers of the Centre National de la Recherche Scientifique (CNRS) of France, and the University of Hawaii. This work is based in part on data products produced at Terapix available at the Canadian Astronomy Data Centre as part of the Canada-France-Hawaii Telescope Legacy Survey, a collaborative project of NRC and CNRS. Partly based on observations obtained as part of the VISTA Hemisphere Survey, ESO Program, 179.A-2010 (PI: McMahon) and of the VST ATLAS Survey, ESO Program, 177.A-­3011 \citep{2015MNRAS.451.4238S}. This paper makes use of data obtained from the Isaac Newton Group Archive which is maintained as part of the CASU Astronomical Data Centre at the Institute of Astronomy, Cambridge. Based on observations made with ESO Telescopes at the La Silla Paranal Observatory under programme ID 084.A-9001. We acknowledge Edward L. Wright and James Schombert for writing the cosmology calculator used throughout this paper. We have extensively used the NumPy \citep{numpy}, SciPy \citep{scipy}, Matplotlib \citep{matplotlib} and AstroPy \citep{astropy} packages. This research made use of Montage, funded by the National Aeronautics and Space Administration's Earth Science Technology Office, Computational Technologies Project, under Cooperative Agreement Number NCC5-626 between NASA and the California Institute of Technology. The code is maintained by the NASA/IPAC Infrared Science Archive. This research has made use of the NASA/IPAC Extragalactic Database (NED) which is operated by the Jet Propulsion Laboratory, California Institute of Technology, under contract with the National Aeronautics and Space Administration. This research has made use of NASA's Astrophysics Data System.  This research has made use of the VizieR catalogue access tool, CDS, Strasbourg, France. The original description of the VizieR service was published in \citet{2000A&AS..143...23O}. This research has made use of ``Aladin sky atlas" developed at CDS, Strasbourg Observatory, France \citep{2000A&AS..143...33B,2014ASPC..485..277B}.

\bibliographystyle{mn2e.bst}

\bibliography{Halpha_cluster_survey}
\appendix
\section{Cluster properties}\label{appendix}

Below, we describe each cluster from our sample in detail. As in Table~\ref{tab:clusters}, the targets are separated in relaxed and merging, and presented in increasing redshift order.

\subsection{Relaxed}

\subsubsection{A1689}

A1689 is an X-ray bright \citep{2007A&A...469..363B}, relaxed, massive ($M_\mathrm{200}$=$2.0^{+0.5}_{-0.3}\times10^{14}M_{\sun}$), strong-lensing cluster at $z=0.183$, which hosts the largest known Einstein partial ring \citep{2010ApJ...723.1678C}. The relaxed nature of the cluster is also supported by spectroscopic data, which indicates the cluster is concentrated, with minimal infall onto the cluster \citep{2009ApJ...701.1336L}. \citet{2002MNRAS.335...10B} performed a spectroscopic H$\alpha$ analysis and found that relative to the field and after accounting for the different spiral fraction, the cluster H$\alpha$ luminosity function is lower by $\sim50$ per cent.

\subsubsection{A963}
Relaxed cluster A963 ($z=0.206$) has an almost perfect Einstein ring around its brightest cluster galaxy \citep[BCG;][]{1991MNRAS.249..184E}. The cluster has a weak lensing mass of $M_\mathrm{200}=7.6^{+1.5}_{-1.3}\times10^{14} M_{\sun}$ \citep{2016MNRAS.461.3794O} and an X-ray luminosity $L_\mathrm{X,0.1-2.4keV}\sim6\times10^{44}$\,erg\,s$^{-1}$ \citep{2010PASJ...62..811O}. \citet{2007ApJ...668L...9V} did pioneering HI work, detecting neutral hydrogen in field galaxies and blue galaxies at the cluster outskirts, however not having any detection for the counterparts located at the cluster core. \mbox{\citet{1994MNRAS.268..393D}} found that the cluster hosts an high number of dwarf galaxies compared to the field.

\subsubsection{A1423}
The relaxed cluster A1423 ($z=0.213$) has a low weak-lensing mass of $M_\mathrm{200}=4.6^{+1.2}_{-1.0}\times10^{14} M_{\sun}$ \citep{2016MNRAS.461.3794O}. As part of the CLASH programme \citep{2012ApJS..199...25P}, the cluster was found to be also be strong-lensing \citep{2015ApJ...801...44Z}. 

\subsubsection{A2261}
The borderline relaxed $z=0.224$ A2261 cluster has a weak lensing mass of $M_\mathrm{200}=12.75^{+2.3}_{-1.5}\times10^{14} M_{\sun}$ \citep{2016MNRAS.461.3794O}. The cluster hosts one of the largest BCGs known. \citet{2012ApJ...757...22C} suggest that the cluster was formed at $1.7<z<2.9$.

\subsubsection{A2390}
A2390 ($z=0.228$) is a relaxed cluster with a weak lensing mass of $M_\mathrm{200}=11.1^{+1.9}_{-1.7}\times10^{14} M_{\sun}$ \citep{2016MNRAS.461.3794O}. This X-ray luminous cluster \citep[$L_\mathrm{X,0.1-2.4keV}\sim12.7\times10^{44}$\,erg\,s$^{-1}$,][]{2010PASJ...62..811O}, hosts diffuse radio emission with irregular morphology (sharp edges towards south and east and filaments towards the north), associated with sloshing around the central, dominant galaxy \citep[mini-halo][]{2003A&A...400..465B}. \citet{1996ApJ...471..694A} concluded that only $5$ per cent of the cluster members have SF at levels higher than typical spirals, indicating that the cluster has been accreting field galaxies for $>8$\,Gyr whose SF has been promptly truncated in the infall process. 

\subsubsection{Z2089}
Z2089 is a relaxed cluster at $z=0.2343$ with $L_\mathrm{X,0.1-2.4keV}\sim6.8\times10^{44}$\,erg\,s$^{-1}$ \citep{1998MNRAS.301..881E} and weak lensing mass $M_\mathrm{200}\sim5\times10^{14} M_{\sun}$  \citep{2006ApJ...653..954D}. The cluster has a prominent central red source, which possibly hosts dusty AGN \citep{2008ApJS..176...39Q}.

\subsubsection{RXJ2129}
RXJ2129.6+0005 (RXJ2129) is a relatively bright, relaxed cluster at $z=0.235$  \citep{2004A&A...425..367B}, which hosts a mini-halo around the radio source at the centre of the cluster \citep{2015A&A...579A..92K}. The cluster X-rays are elongated in the NW-SW direction \citep{2010ApJ...719.1619O, 2015A&A...579A..92K}. The cluster has a weak lensing mass of $M_\mathrm{200}=5.3^{+1.8}_{-1.4}\times10^{14} M_{\sun}$ \citep{2010PASJ...62..811O}.

\subsubsection{RXJ0437}
RX J0437.1+0043 (RXJ0437) is a relaxed cluster \citep[$z=0.285$,][]{2004A&A...425..367B} with a weak lensing mass $M_\mathrm{200}\sim5\times10^{14} M_{\sun}$ \citep{2006ApJ...653..954D}. The elliptical X-ray morphology is consistent with a relaxed state \citep{2005A&A...444..157F}.

\subsection{Merging}

\subsubsection{A545}
A545 is at a redshift of $0.154$ and has an X-ray luminosity of $L_\mathrm{X,0.1-2.4keV}\sim5.05\times10^{44}$\,erg\,s$^{-1}$ \citep{2004A&A...425..367B}. \citet{2011A&A...529A.128B} performed a detailed spectroscopic and X-ray analysis of the cluster and find an extremely complex and disturbed morphology with at least three subclusters and no dominant galaxy. At the centre of the cluster, there exists a `star pile', an extended low-surface brightness feature with three nuclei, which \citet{2011A&A...528A..61S} interpret as the remnant of a tidally stripped galaxy or galaxies. \citet{2004A&A...425..367B} find evidence for an X-ray shock coinciding with the northern edge of the regular, centrally located radio halo \citep[which was studied in detail by][]{2003A&A...400..465B}. \citet{2011A&A...529A.128B} interpreted their data as indicative of a merger happening in two directions, within the plane of the sky. Based on their spectra, \citet{2011A&A...529A.128B} calculated a mass of about $M\sim (11-18)\times 10^{14} M_{\sun}$.

\subsubsection{A3411}
Based on X-ray data, A3411 \citep[$z=0.169$;][]{2002ApJ...580..774E} is a complex merging cluster which is possibly interacting with the nearby A3412 and hosts both a radio halo and a $1.9$-Mpc radio relic towards the south-east of the cluster \citep{2013MNRAS.435..518G, 2013ApJ...769..101V}. X-ray and radio data indicate that the relic is possibly formed by a weak shock ($M<1.3$) re-accelerating fossil plasma from a nearby radio AGN \citep{weeren2016}. The emerging scenario is that of a binary $1:1$ merger, happening in the plane of the sky in the NW-SE direction and observed 1\,Gyr after core passage \citep{weeren2016}. The northern subcluster (A3411) survived the collision, while the southern subcluster (A3412) was stripped of its gas during the merger \citep{weeren2016}. Based on a dynamical analysis, \citet{weeren2016} estimate a mass of $14^{+4}_{-3}\times10^{14} M_{\sun}$ and $18^{+5}_{-4}\times10^{14} M_{\sun}$ for A3411 and A3412, respectively.

\subsubsection{A2254}
Based on an optical and X-ray analysis, \citet{2011A&A...536A..89G} classify A2254  \citep[$z=0.178$;][]{2001A&A...376..803G} as a binary, post merger cluster, which hosts a radio halo \citep{1999NewA....4..141G}. Based on spectroscopy, \citet{2011A&A...536A..89G} estimate the total mass of the system to be about $(15-29)\times10^{14} M_{\sun}$. The relative line-of-sight (LOS) velocity of $\sim3000$\,km\,s$^{-1}$ and the projected linear distance between the two subclusters of $\sim 0.5$\,Mpc are consistent with a young merger, with core passage happening $<0.5$\,Gyr ago.

\subsubsection{CIZA J2242.8+5301}
CIZA J2242.8+5301 \citep[`Sausage', $z=0.188$;][]{2007ApJ...662..224K, 2015ApJ...805..143D} is a merging galaxy cluster hosting double, symmetric radio-detected shocks perpendicular to the merger axis \citep{2010Sci...330..347V}. Radio modelling and X-ray data indicate a Mach number $M\sim3$ for the main $1.4$-Mpc shock \citep{2013PASJ...65...16A, 2014MNRAS.440.3416O, 2014MNRAS.438.1377S}, however some studies find a higher Mach number of $\sim4.5$ \citep{2010Sci...330..347V, 2016MNRAS.462.2014D}. The cluster is consistent with a massive post-core passage merger between two clusters of similar masses $M_\mathrm{200}=11.0^{+3.7}_{-3.2}\times10^{14}M_{\sun}$ and
 $9.8^{+3.8}_{-2.5}\times10^{14}M_{\sun}$ \citep[weak lensing analysis consistent with dynamical analysis;][]{2015ApJ...802...46J,2015ApJ...805..143D}, with the merger happening about $0.5-1.0$\,Gyr ago \citep{2011MNRAS.418..230V, 2014MNRAS.445.1213S}. The cluster was found to host a significant overdensity of H$\alpha$ emitters, which are more massive, more HI gas rich and more SF than their field counterparts \citep{2014MNRAS.438.1377S, Stroe2015, 2015MNRAS.452.2731S} and have evidence for outflows from supernovae and AGN activity \citep[from spectroscopy,][]{2015MNRAS.450..630S}.

\subsubsection{A115}
\citet{1981ApJ...243L.133F} found that A115 at $z=0.1971$ has a double X-ray peak, consistent with two subclusters with substantial off-axis motion \citep{2005ApJ...619..161G}. The X-ray luminosity of the cluster is $L_\mathrm{X,0.1-2.4keV}\sim9\times10^{44}$\,erg\,s$^{-1}$, while its weak lensing mass is $6.7^{+3.2}_{-2.1}\times10^{14} M_{\sun}$ \citep{2016MNRAS.461.3794O}. \mbox{\citet{2007A&A...469..861B}} performed a spectroscopic study of A115 and found that the galaxies in the northern, less massive subcluster are experiencing higher SF activity compared to the southern subcluster. They propose a pre-merging scenario where the two subclusters are colliding at a LOS velocity of $1600$\,km\,s$^{-1}$ and will cross within 0.1\,Gyr. However this scenario is not fully consistent with the presence of arc-like diffuse emission extended over 2\,Mpc \citep{2001A&A...376..803G}, cospatial with a $M\sim1.8$ X-ray shock \citep{2016MNRAS.460L..84B}, which indicates the presence of a merger shock perpendicular to the merger axis.

\subsubsection{A2163}
A2163 is an exceptionally hot, luminous, massive ($M_\mathrm{200} = 29.0^{+4.6}_{-5.8}\times10^{14} M_{\sun}$) merging cluster at $z=0.203$ \citep{2001A&A...373..106F,2011ApJ...741..116O}. The optical analysis performed by \citet{2008A&A...481..593M} reveals a complex merging scenario: the cluster has a main bi-modal central component, a northern component as well as two other substructures. \citet{2008A&A...481..593M} infer the main clump has undergone a recent merger in the last 0.5\,Gyr along the NW-SW direction, probably with a non-zero impact parameter \citep{2011ApJ...741..116O}, with the northern component infalling into the cluster. A weak lensing analysis indicated that the two main clump components have a mass ratio of $1:8$ \citep{2011ApJ...741..116O}. \citet{2011ApJ...741..116O} also found an offset between the X-ray distribution and the galaxy density, attributed to ram pressure stripping of gas away from the dark matter host. The cluster also hosts a giant radio halo, indicative of increased turbulence in the main clump \citep{2001A&A...373..106F}

\subsubsection{A773}
A773 ($z=0.217$) is a binary  merging cluster with $\sim4:1$ mass ratio, merging in the NE-SW direction, with a weak lensing mass of $M_\mathrm{200}=10.2^{+1.5}_{-1.3}\times10^{14} M_{\sun}$ \citep{2016MNRAS.461.3794O}. \citet{2004ApJ...605..695G} found that one of the two galaxy subclumps coincides with the centre of the X-ray emission, while a radio halo is located in the cool region between the two subclusters. \citet{2007A&A...467...37B}, using spectroscopic data, concluded the cluster is in an advanced stage of merging with an impact velocity of $\sim2300$\,km\,s$^{-1}$.

\subsubsection{1RXS J0603.3+4214}
1RXS J0603.3+4214 (`Toothbrush' cluster, $z=0.225$) was discovered as a merging cluster with diffuse radio emission in the form of at least one 1.9-Mpc, linear shock perpendicular to the merger axis and a halo by \citet{2012A&A...546A.124V}. Radio observations indicate a Mach number of $\sim2.8$, in tension with X-ray observations which predict a much lower value of $\sim1.2$ \citep{2016ApJ...818..204V}. The merger scenario is complicated, with two massive colliding clumps with a mass ratio of $3:1$ ($M_\mathrm{200}=6.3^{+2.2}_{-1.6}\times10^{14}M_{\sun}$ and $M_\mathrm{200}=2.0^{+1.2}_{-0.7}\times10^{14}M_{\sun}$), as well as 1-2 smaller clumps participating in the merger \citep[weak lensing analysis,][]{2016ApJ...817..179J}. This view is roughly consistent with hydrodynamical simulations by \citet{2012MNRAS.425L..76B} who also find a complicated merger scenario necessary and place the main clump core passage at about $2$\,Gyr ago. The cluster was found to have a similar density of H$\alpha$ emitters as field galaxies at the same redshift \citep{2014MNRAS.438.1377S, Stroe2015}.

\subsubsection{A2219}
A2219 \citep{2010PASJ...62..811O} is one of the hottest, most X-ray luminous clusters known \citep{2015arXiv150505790C}. A post-merger cluster at $z=0.2256$ with weak lensing mass of $M_\mathrm{200}=10.9^{+2.2}_{-1.8}\times10^{14} M_{\sun}$ \citep{2016MNRAS.461.3794O}, A2219 hosts a radio halo with regular and symmetric structure \citep{2003A&A...400..465B}. \citet{2004A&A...416..839B} performed a detailed spectroscopic study of A2219 and found a high velocity dispersion, from which they derive a total mass of $\sim28\times 10^{14} M_{\sun}$. Their data indicate a merger axis on the SE-NW direction, inclined at about $45^{\circ}$ from the plane of the sky \citet{2004A&A...416..839B}. \citet{2015arXiv150505790C} find two shocks and a cold front perpendicular on the merger axis, coincident with the edges of the radio halo, and estimate a times since core passage of about $\sim0.26$\,Gyr. 

\subsubsection{A1300}
A1300 is a hot, luminous \citep{2004A&A...425..367B} post-merger cluster at $z=0.3072$ \citep{1997A&AS..124..283P, 2012MNRAS.420.2480Z}. The cluster hosts a halo \citep{1999MNRAS.302..571R}, and has evidence for a $M=1.2$ shock from X-ray data \citep{2012MNRAS.420.2480Z} coincident with a radio relic towards the south-west edge. Comparison with simulations indicates that $\sim3$\,Gyr have passed since core passage, to form a system with $M_\mathrm{200}\sim6\times10^{14} M_{\sun}$ \citep{2012MNRAS.420.2480Z}.

\subsubsection{A2744}
A2744 ($z=0.308$) is an extremely disturbed, complex and young merging cluster with high X-ray luminosity \citep{2001A&A...376..803G} and large weak lensing mass $M_\mathrm{200}=20.6\pm4.2\times10^{14} M_{\sun}$ \citep{2016ApJ...817...24M}. The cluster hosts at least 4 substructures with mass ratios approximately $2:1:1:1$. \citet{2011MNRAS.417..333M} propose a scenario of a simultaneous double merger happening $0.12-0.15$\,Gyr ago, one bullet-like merger in the NE-SW and the other in the NW-SE direction. \citet{2012ApJ...750L..23O} find galaxies with trails of SF which are affiliated with the Bullet-like subcluster and the X-ray shock.

\end{document}